\documentclass[final,authoryear,3p,times]{elsarticle}

\usepackage{afterpage}
\usepackage{algorithm}
\usepackage{algpseudocode}
\usepackage{amsmath}
\usepackage{amssymb}
\usepackage{amsthm}
\usepackage{bm}
\usepackage{booktabs}
\usepackage{calc}
\usepackage{caption}
\usepackage{subcaption}
\usepackage[table]{xcolor} 
\usepackage{enumerate}
\usepackage{graphicx}
\usepackage{ifthen}
\usepackage[parfill]{parskip} 
\usepackage{multicol}
\usepackage{siunitx}
\usepackage{ulem}

\usepackage[colorlinks]{hyperref}
\usepackage[nameinlink,capitalise]{cleveref}
\crefname{appendix}{}{}

\usepackage{tikz}
\usetikzlibrary{arrows}
\usetikzlibrary{automata}
\usetikzlibrary{calc}
\usetikzlibrary{chains}
\usetikzlibrary{positioning}
\usetikzlibrary{shapes}

\usepackage{varwidth}

\usepackage{tcolorbox}
\newtcolorbox[auto counter]{problem}[2][]{colframe=blue!30, colback=blue!5, coltitle=black, title=Problem~\thetcbcounter~ #2,#1}

\usepackage{spverbatim} 
\usepackage{listings} 

\definecolor{Demiray}{rgb}{0.1725, 0.6275, 0.1725}
\definecolor{Gent}{rgb}{1.0, 0.4902, 0.0431}
\definecolor{Holzapfel}{rgb}{0.1216, 0.4667, 0.7059}
\definecolor{MooneyRivlin}{rgb}{0.8392, 0.1529, 0.1569}
\definecolor{Ogden}{rgb}{0.5804, 0.4039, 0.7412}
\definecolor{BlatzKo}{rgb}{0.5490, 0.3373, 0.2941}
\definecolor{NeoHooke}{rgb}{0.7373, 0.7412, 0.1333}

\edef\svtheparindent{\the\parindent}
\usepackage{parskip}
\parindent=\svtheparindent\relax

\newcommand{\MF}{Material Fingerprinting}

\definecolor{ForestGreen}{RGB}{34,139,34}
\definecolor{InternationalOrange}{rgb}{1.0, 0.31, 0.0}
\definecolor{WineRed}{RGB}{139,0,0}

\newcommand{\CHANGE}[1]{\textcolor{black}{#1}}
\newcommand{\CHANGETWO}[1]{\textcolor{black}{#1}}

\newcommand{\boldface}[1]{\boldsymbol{#1}}  

\newcommand{\bff}{\boldface{f}}

\newcommand{\bfC}{\boldface{C}}

\newcommand{\bfF}{\boldface{F}}

\newcommand{\bfP}{\boldface{P}}

\newcommand{\bfalpha}{\boldsymbol{\alpha}}

\newcommand{\bftheta}{\boldsymbol{\theta}}

\newcommand{\Rset}{\mathbb{R}}

\DeclareMathOperator{\tr}{tr}

\newcommand{\argmax}{\operatornamewithlimits{arg\ max}}

\newcommand{\be}{\begin{equation}}
\newcommand{\ee}{\end{equation}}
\newcommand{\bea}{\begin{equation}\begin{aligned}}
\newcommand{\eea}{\end{aligned}\end{equation}}
\newcommand{\beq}{\begin{eqnarray}}
\newcommand{\eeq}{\end{eqnarray}}
\newcommand{\bem}{\begin{multline}}
\newcommand{\eem}{\end{multline}}
\newcommand{\ba}{\begin{align}}
\newcommand{\ea}{\end{align}}
\newcommand{\bcase}{\left\{ \begin{array}{ll}}
\newcommand{\ecase}{\end{array} \right.}
\begin{document}

\begin{frontmatter}

\title{Material Fingerprinting: \\ A shortcut to material model discovery without solving optimization problems}

\author[fau,stan]{Moritz Flaschel\corref{cor1}}
\author[fau,stan]{Denisa Martonová}
\author[stan]{Carina Veil}
\author[fau,stan]{Ellen Kuhl}

\cortext[cor1]{Correspondence: moritz.flaschel@fau.de}

\address[fau]{Institute of Applied Mechanics, Egerlandstraße 5, Friedrich-Alexander-Universität Erlangen–Nürnberg, 91058 Erlangen, Germany}
\address[stan]{Department of Mechanical Engineering, Stanford University, 440 Escondido Mall, California 94305, United States.}

\begin{abstract}

We propose Material Fingerprinting, a new method for the rapid discovery of mechanical material models from direct or indirect data that avoids solving potentially non-convex optimization problems. The core assumption of Material Fingerprinting is that each material exhibits a unique response when subjected to a standardized experimental setup. We can interpret this response as the material’s fingerprint, essentially a unique identifier that encodes all pertinent information about the material’s mechanical characteristics. Consequently, once we have established a database containing fingerprints and their corresponding mechanical models during an offline phase, we can rapidly characterize an unseen material in an online phase. This is accomplished by measuring its fingerprint and employing a pattern recognition algorithm to identify the best matching fingerprint in the database. In our study, we explore this concept in the context of hyperelastic materials, demonstrating the applicability of Material Fingerprinting across different experimental setups. Initially, we examine Material Fingerprinting through experiments involving homogeneous deformation fields, which provide direct strain-stress data pairs. We then extend this concept to experiments involving complexly shaped specimens with heterogeneous deformation fields, which provide indirect displacement and reaction force measurements. Focusing on numerically generated data in this initial study, we show that, in both experimental setups, Material Fingerprinting is an efficient tool for model discovery, bypassing the challenges of potentially non-convex optimization. While investigating isotropic hyperelasticity in this work, we believe that Material Fingerprinting provides a powerful and generalizable framework for rapid material model identification across a wide range of experimental designs and material behaviors, paving the way for numerous future developments.

\end{abstract}


\end{frontmatter}





\begin{figure}[h!]
\centering
\includegraphics[width=\linewidth]{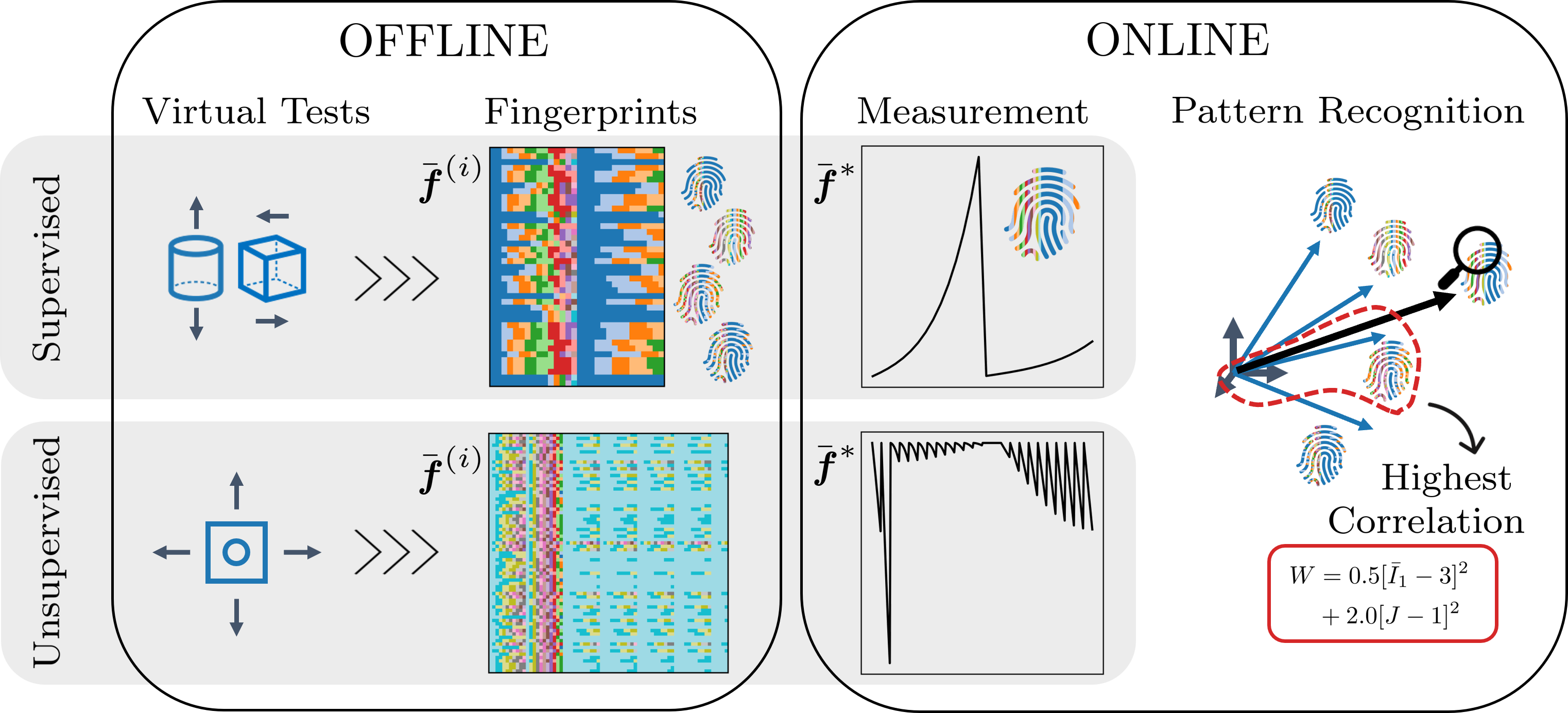}
\caption{Schematic overview of the \MF{} workflow. The top row depicts the supervised approach using tests with homogeneous deformation fields, while the bottom row illustrates the unsupervised case based on heterogeneous deformations. In the offline phase, a database of material fingerprints is generated synthetically. During the online phase, the fingerprint of an unseen material is measured and matched to the closest entry in the database using a pattern recognition algorithm, enabling fast material model discovery.}
\label{fig:abstract}
\end{figure}

\section{Introduction}
\label{sec:introduction}
To understand the physical behavior of materials and accurately simulate their mechanical behavior under complex shapes and loading conditions, it is essential to develop mathematical models that accurately describe the materials' mechanics.
This mechanical characterization of materials traditionally involves solving an optimization problem.
In this scenario, 
either the material parameters of a phenomenological model \citep{avril_overview_2008,pierron_towards_2020} or the weights and biases of a machine learning-based material model \citep{fuhg_review_2024} are adjusted to minimize the discrepancy between model predictions and observed data.
However, these \textbf{optimization processes encounter significant challenges}.
Firstly, depending on the number of parameters and the complexity of the model, solving the optimization problem is time-consuming.
This issue becomes more severe when the evaluation of the objective function requires solving ordinary or partial differential equations, as is often the case in material characterization from full-field measurements \citep{avril_overview_2008,roux_optimal_2020,romer_reduced_2024}.
Secondly, optimization problems in mechanical material characterization often feature non-convex objective functions with multiple local optima.
This issue is particularly pronounced in machine learning-based models, where the loss function landscape is highly non-convex.
Thirdly, imposing physical constraints on the material behavior can introduce nonlinear parameter constraints \citep{xu_discovering_2025} within the optimization problem, thereby complicating its solution.

In response to these challenges, we propose a database approach that completely eliminates the need for solving optimization problems by utilizing an efficient pattern recognition algorithm.
We name this method \MF{}, drawing inspiration from the pioneering Magnetic Resonance Fingerprinting technique \citep{ma_magnetic_2013}. 
The concept of \MF{} is based on the notion that, within a given experimental setup, \textbf{each material possesses a unique fingerprint that characterizes its mechanical behavior}.
\MF{} involves a two-stage procedure as graphically illustrated in \cref{fig:abstract}:
During the offline stage, we create a comprehensive database of characteristic material fingerprints.
Then, in the online stage, we employ a pattern recognition algorithm to identify an appropriate material model for a given experimental dataset by referencing this pre-established database.
This strategy offers several distinct advantages:
\begin{itemize}
    \item \textbf{Model discovery}: By considering multiple models during database generation, \MF{} enables the discovery of models. \CHANGE{Here, we use the term model discovery as opposed to traditional model calibration.} Unlike merely calibrating parameters of an a priori selected model, \MF{} simultaneously \CHANGE{discovers} the optimal functional form \CHANGE{from a predefined set of functional forms} and its parameters to accurately describe material behavior.
    \item \textbf{No optimization}: \MF{} eliminates the need for solving complex optimization problems. Once the fingerprint database is generated in an offline phase, it can be repeatedly used for material model discovery in the online phase, which involves a pattern recognition algorithm that is computationally efficient and inherently amenable to parallelization.
    \item \textbf{No local optima}: Traditional material model discovery approaches that rely on optimization can be trapped in local optima. \MF{} circumvents this issue, as its pattern recognition algorithm identifies the best matching fingerprint from the database, which can be interpreted as a discrete global optimum within the searchable database.
    \item \textbf{Applicability to any experiment}: The concept of \MF{} is versatile and can be applied to any experimental setup. This work demonstrates its applicability to both homogeneous and heterogeneous tests, utilizing either direct or indirect data.
    \item \textbf{Applicability to various material behaviors}: \MF{} is not limited to specific types of material models. Although this study focuses on hyperelasticity, \CHANGE{we anticipate that various} material models can be incorporated during database generation. \CHANGE{Alternatively, multiple databases covering different material classes, such as elasticity, viscoelasticity, and plasticity, may be generated.}
    \item \textbf{Physical admissibility}: 
    During the creation of the fingerprint database, the selection of material models and parameter ranges is fully controlled. Consequently, \MF{} ensures the discovery of only physically admissible material models.
    \item \textbf{Interpretability}: Material models discovered through \MF{} exhibit a small number of material parameters and physically interpretable functional forms. The interpretability of these models can be further enhanced by introducing sparsity promotion into the pattern recognition algorithm.
\end{itemize}

Before explaining our proposed \MF{} in detail, we first provide a summary of various methods for mechanical material characterization, covering both traditional techniques and recent machine learning-based approaches. We then discuss methods related to our \MF{} that rely on databases or lookup tables.

The traditional strategy for characterizing the mechanical behavior of materials is to assume a mathematical model that describes the material behavior, and to calibrate the corresponding material parameters in this model such that the model response is in good agreement with the experimentally acquired data \citep{lemaitre_mechanics_1994,hartmann_parameter_2001}.
In recent years, however, it was realized that fixing a traditional model and calibrating its parameters is often not flexible enough to capture the complex mechanical behavior observed during experiments.
This motivated the development of more modern data-driven and machine learning-based approaches for mechanical material characterization \citep{fuhg_review_2024}.
For instance, the mechanical response of a material may be described by a black-box machine learning model.
Prominent approaches are based on neural networks \citep{ghaboussi_knowledgebased_1991,mozaffar_deep_2019,masi_thermodynamics-based_2021,huang_variational_2022,asad_mechanics-informed_2022,klein_polyconvex_2022,rosenkranz_comparative_2023,flaschel_convex_2025}, splines \citep{latorre_what-you-prescribe-is-what-you-get_2014,wiesheier_versatile_2024}, Gaussian processes \citep{frankel_tensor_2020,fuhg_local_2022}, neural ordinary differential equations \citep{tac_data-driven_2022,jones_attention-based_2025}, or parametrized non-smooth convex sets \citep{bleyer_learning_2025}.
Another line of research avoids the explicit formulation of a material model by running simulations directly informed by data \citep{kirchdoerfer_data-driven_2016,ibanez_manifold_2018}.
Further, recognizing that both machine learning-based and model-free approaches lack physical interpretability, a different direction focuses on the automated discovery of interpretable material models from data using symbolic or sparse regression \citep{schoenauer_evolutionary_1996,ratle_grammar-guided_2001,versino_data_2017,flaschel_unsupervised_2021,flaschel_automated_2023-1,flaschel_automated_2023-2,linka_new_2023,linka_automated_2023,marino_automated_2023,holthusen_theory_2024,peirlinck_automated_2024,martonova_automated_2024,martonova_discovering_2025,martonova_generalized_2025,abdolazizi_constitutive_2025,shojaee_automated_2025}.
Unlike traditional methods that calibrate parameters of predefined models, these approaches aim to discover the mathematical structure of the material model itself, while simultaneously identifying its parameters.

An important consideration in characterizing the mechanical behavior of materials is the selection of data used to inform the material model.
A straightforward approach to material model calibration involves experiments that produce homogeneous deformation fields \citep{lemaitre_mechanics_1994,hartmann_parameter_2001}.
These experiments have the advantage of providing stress–strain data pairs that can be directly used to calibrate material models.
However, they are typically limited to specific deformation modes, such as uniaxial tension or simple shear, which may adversely affect the model’s predictive accuracy under more general loading conditions.
Recognizing this issue, emerging methods, like Finite Element Model Updating \citep{kavanagh_finite_1971,kavanagh_extension_1972}, the Virtual Fields Method \citep{grediac_principle_1989,pierron_virtual_2012}, or the Equilibrium Gap Method \citep{claire_finite_2004}, calibrate material models based on experiments with heterogeneous deformation fields \citep{avril_overview_2008,roux_optimal_2020,romer_reduced_2024}.
This idea has been extended to learning black-box models \citep{thakolkaran_nn-euclid_2022,benady_unsupervised_2024,shi_deep_2025,moreno-mateos_biaxial_2025,bourdyot_learning_2025}, discovering interpretable models \citep{wang_inference_2021,flaschel_unsupervised_2021}, and characterizing materials using physics-informed neural networks \citep{anton_physics-informed_2022}.

As mentioned earlier, all previously discussed methods for calibrating, learning, or discovering material models rely on optimization problems, which often face challenges such as long computation times and the presence of multiple local minima due to non-convex objective functions.
To address these issues, we propose \MF{}, a database-driven approach for rapid material model discovery.
Database methods are well established in other areas of research: 
For example, in biomedical imaging, magnetic resonance fingerprinting has been proposed to identify physical parameters that influence the magnetic response of different tissue types during magnetic resonance imaging \citep{ma_magnetic_2013,mcgivney_svd_2014}.
In cell biophysics, cell stiffnesses have been identified from 
a lookup table during real-time deformability cytometry \citep{wittwer_new_2023}.
In rheology, viscoelastic material model parameters have been calibrated from a database using group shear wave speed measurements \citep{rouze_characterization_2018, trutna_robust_2019, trutna_viscoelastic_2020,trutna_measurement_2020}.
Surprisingly, the potential of \MF{} has not yet been explored in the context of mechanical material model discovery.
A key distinction of \MF{} is that it does not assume a predefined functional form for the material model; instead, it discovers the model structure automatically as part of the inverse problem. Furthermore, \MF{} represents the first database-based approach applied to mechanical characterization in the context of heterogeneous deformation fields.

We note that the term \MF{} appears in the literature in contexts unrelated to mechanical material modeling and should be distinguished from our proposed method,
for instance, 
in the context of 
extracting structural information from atom probe microscopy data
\citep{spannaus_materials_2021},
in the context of 
spectral fingerprints to compare the similarity of different materials based on their electronic properties
\citep{kuban_similarity_2022},
in the context of material surfaces 
\citep{filip_material_2024}, or 
in the context of 
feature compression and graphical representation of single crystal materials
\citep{jaafreh_introducing_2025}.
Although these methods share the name \MF{} with our approach, they do not involve the discovery of mechanical material models and are therefore conceptually different.


\section{\MF{}}
\label{sec:material_fingerprinting}

\MF{} involves two stages, an offline stage that constructs a database of fingerprints and an online stage that uses a pattern recognition algorithm to discover a material model for an unseen experimental measurement.
Before describing the workflow of \MF{} in detail, we provide a brief explanation of each stage of the process:
\begin{itemize}
    \item \textbf{Generating the fingerprint database (offline)}: To generate a database of material fingerprints, one or multiple virtual experiments are conducted for a variety of different material models and material parameters. A key assumption during the database generation is that the experiments follow a standardized design. This means that the experiments should be designed such that they can be reproduced for new materials in different laboratories. The experiments may be as simple as uniaxial tension tests that deliver stress-strain data pairs, or more complex such as full-field displacement and net force measurements of complexly-shaped specimens. Specifically in the latter case, the specimen geometry should be standardized in a way that it can be reproduced for different materials.
    
    For each of the considered experiments, the nature of the corresponding fingerprint must be defined. For example, for a uniaxial tension experiment of a homogeneous specimen, the fingerprint may be a vector of stress measurements for a predefined set of applied stretches. For a more complex experiment with a heterogeneous deformation field, the fingerprint may contain both net force measurements as well as displacement measurements at a predefined representative set of locations on the specimen surface. After generating the fingerprints for various material models and different combinations of parameters, the fingerprints are normalized and stored, for example, in a cloud-based shared database together with the corresponding information about the material models and material parameters. Once the database is generated, it can be used repeatedly to characterize new, unseen materials.
    
    \item \textbf{Pattern recognition algorithm (online)}: Once the fingerprint database is generated and made publicly accessible, for example, via the Internet, it can be used for automated material model discovery. The user performs the standardized experiments as defined during the offline stage, but now for a new material, and the material’s fingerprint is measured. Afterwards, a pattern recognition procedure is used to search the fingerprint database for the most similar fingerprint and the corresponding material model and its parameters are extracted. This step is computationally inexpensive and requires less computational resources and time than solving a potentially non-convex optimization problem. This search is also easily parallelizable.
     
\end{itemize}

In this work, we distinguish between two types of experiments:
First, we consider mechanical experiments of samples that exhibit a simple \textbf{homogeneous} deformation field.
Such experiments deliver labeled strain-stress data pairs.
As these data pairs establish a direct connection between strains and stress, they are often referred to as \textbf{direct data}, and methods that use these data are denoted as \textbf{supervised}.
Second, we consider mechanical experiments of complexly-shaped specimens resulting in \textbf{heterogeneous} deformation fields.
In these experiments, the displacement field in the interior of the specimen is measured using full-field measurement equipment such as Digital Image Correlation or Digital Volume Correlation, while net reaction forces are measured at the boundary of the specimen.
Such experiments have the advantage that, in a single experiment, a wide range of different deformation states are excited in the material, facilitating the three-dimensional mechanical characterization of a material from a single experiment \citep{grediac_principle_1989,hild_digital_2006,avril_overview_2008,pierron_towards_2020}.
Because these experiments do not deliver labeled strain-stress data pairs, their resulting data are referred to as \textbf{indirect data}, and methods that are informed by such data are called \textbf{unsupervised} in machine learning jargon \citep{flaschel_unsupervised_2021}.
We note that, in essence, the supervised and unsupervised approaches to \MF{} differ primarily in how the fingerprints are defined and the databases are generated.
However, the overarching concept of the method remains consistent across both approaches, see \cref{fig:abstract}.

\begin{figure}
    \centering
    \includegraphics[width=0.8\textwidth]{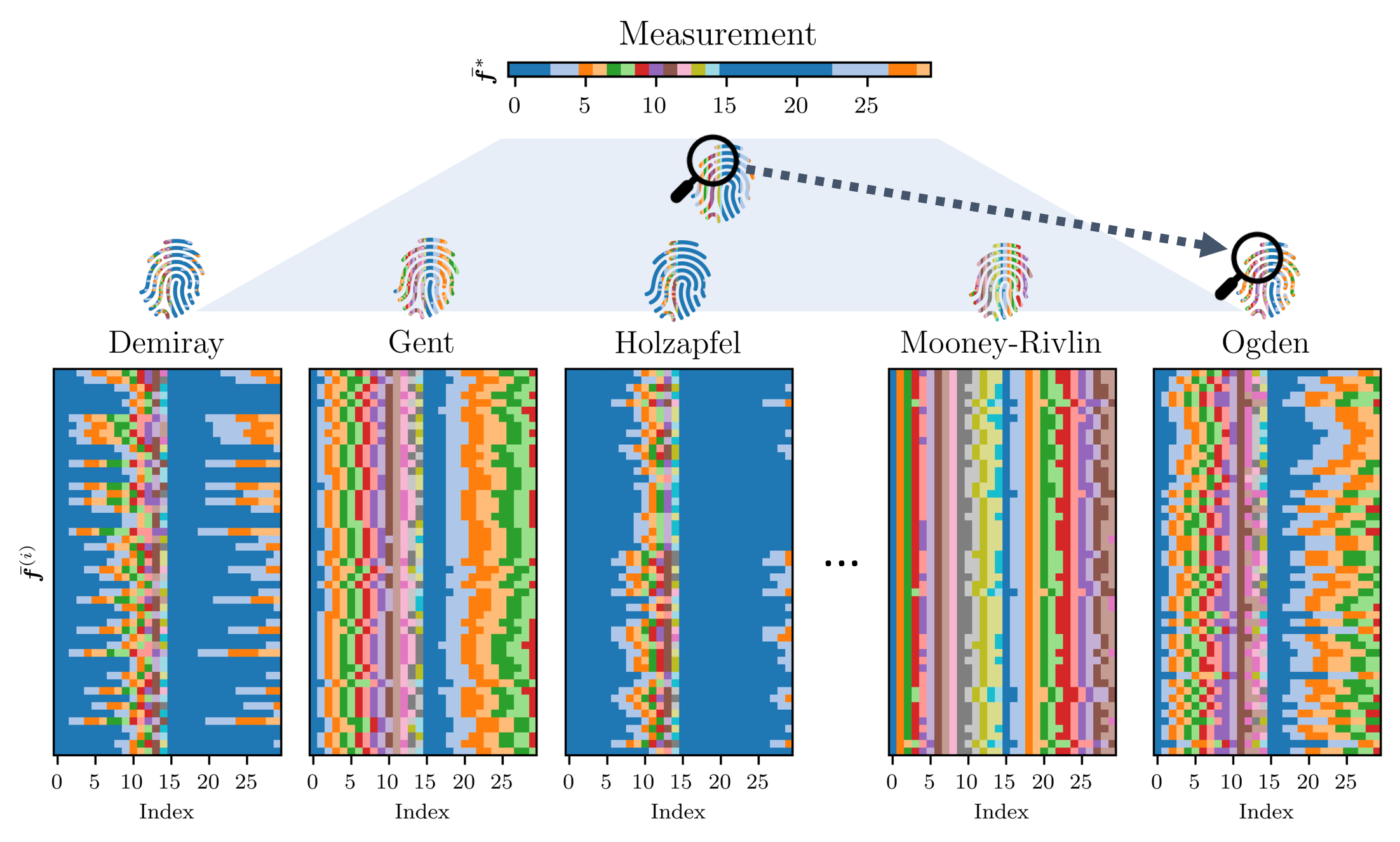}
    \caption{
    Illustration of supervised Material Fingerprinting. The measurement is compared to all fingerprints in the database. Each row in the database represents a fingerprint computed using a specific set of model parameters, with the color qualitatively indicating differences in the magnitude of the fingerprint components. While the full database contains a large number of fingerprints derived from various models, the figure illustrates a subset only. The fingerprints are grouped and displayed by model type, excluding simple models that have a single parameter only.
    }
    \label{fig:pattern_recognition_matrices}
\end{figure}

\subsection{Supervised \MF{}}
We first focus on supervised \MF{} and consider simple experiments with homogeneous deformation fields.
In this work, we focus on a combination of uniaxial tension and simple shear experiments, noting that the method is not limited to these types of experimental setups.
We put our attention on hyperelastic materials in this work, but again note that the method can be easily extended to dissipative materials.
In the following, we detail the experimental design, the definition of the material fingerprints, the creation of the fingerprint databases, the fingerprint normalization, and the pattern recognition algorithm.

\subsubsection{Experiment design}
In uniaxial tension (UT) and simple shear (SS) tests, the deformation gradient is assumed constant in space.
\CHANGE{Assuming incompressible material behavior, it is}
\be
\label{eq: deformation gradient ut ss}
\bfF_{\text{UT}} =
	\begin{bmatrix}
		F_{11} & 0 & 0\\
		0 & {1}/{\sqrt{F_{11}}} & 0\\
		0 & 0 & {1}/{\sqrt{F_{11}}}\\
	\end{bmatrix} 
	\quad \mbox{and} \quad
\bfF_{\text{SS}} =
	\begin{bmatrix}
		1 & F_{12} & 0\\
		0 & 1 & 0\\
		0 & 0 & 1\\
	\end{bmatrix},
\ee
in which the longitudinal stretch $F_{11}$ and the shear component $F_{12}$ are imposed on the material during the experiments.
Specifically, we impose a number of $n_{\text{UT}}$ different longitudinal stretches during the uniaxial tension experiment and a number of $n_{\text{SS}}$ simple shear states during the simple shear experiment.

\subsubsection{Fingerprint definition}
For each prescribed deformation during the uniaxial tension experiment, the normal component of the first Piola-Kirchhoff stress $P_{11}$ is measured, and each imposed simple shear state yields a shear component $P_{12}$ measurement.
Different materials exhibit different stress responses during an experiment.
Thus, \textbf{the stress response of the material can be interpreted as its fingerprint}.
Specifically, we introduce the fingerprint vector $\bff \in \Rset^{n_f}$ with $n_f = n_{\text{UT}} + n_{\text{SS}}$ containing all stresses measured during the experiments.
\cref{fig:pattern_recognition_matrices} presents graphical illustrations of the fingerprints corresponding to various material models, where the fingerprint vectors are represented as color-coded rows.

\subsubsection{Fingerprint database generation}
\label{sec: Fingerprint database generation}
To generate fingerprints for the considered experimental setup, a relationship between the deformation and the stress must be established.
The deformation vs. stress relationship of hyperelastic materials is characterized through the strain energy density $W(\bfF; \bftheta, \bfalpha)$, where $\bfF$ is the deformation gradient and $\bftheta$ and $\bfalpha$ are material parameters.
In this work, we distinguish between the parameters $\bftheta$, which fulfill the homogeneity property
\be
W(\bfF; a \ \bftheta, \bfalpha) = a \ W(\bfF; \bftheta, \bfalpha), \quad \forall a \in \Rset,
\ee
as opposed to the parameters $\bfalpha$ not fulfilling this property.
We will thus call $\bftheta$ the homogeneity parameters and $\bfalpha$ the non-homogeneity parameters.
We note that all hyperelastic material models known from the literature \citep{marckmann_comparison_2006,chagnon_hyperelastic_2015,dal_performance_2021} contain homogeneity parameters, while not all models contain non-homogeneity parameters.
For example, models belonging to the Mooney-Rivlin family contain only homogeneity parameters \citep{rivlin_large_1950}, and Ogden-type models contain both types of parameters \citep{ogden_large_1972}.

The first Piola-Kirchhoff stress for hyperelastic materials computes to
\be
\label{eq:Piola-Kirchhoff}
\bfP(\bfF; \bftheta, \bfalpha) = \frac{\partial W(\bfF; \bftheta, \bfalpha)}{\partial \bfF}.
\ee
This equation establishes a relationship between the prescribed deformation $\bfF$ and the measured stress $\bfP$ dependent on the choice of the strain energy density function and its parameters.
For supervised \MF{}, we focus on isotropic and incompressible material behavior in this work.
Under these assumptions, \cref{eq:Piola-Kirchhoff} establishes a relationship between $F_{11}$ and $P_{11}$ for the uniaxial tension experiments and between $F_{12}$ and $P_{12}$ for the simple shear experiments, as detailed in \cref{sec:uniaxial_tension_simple_shear,sec:uniaxial_tension_simple_shear}.

To construct a database of material fingerprints, we numerically compute the fingerprints $\bff$ for different functional forms of the strain energy density and different choices of $\bftheta$ and $\bfalpha$.
The database then consists of data triplets
\be
(\bff^{(i)},\bftheta^{(i)},\bfalpha^{(i)}),
\ee
with $i=1, \dots ,n_d$, where $n_d$ is the size of the database.
We emphasize that computing the fingerprints for different choices of parameters $\bftheta$ and $\bfalpha$ does not mean that the database contains only one material model.
The parameters $\bftheta$ and $\bfalpha$ are potentially large vectors, and setting different entries in $\bftheta$ and $\bfalpha$ to zero yields different functions in the strain energy density and thus different material models.
Here, we choose the incompressible Blatz-Ko \citep{blatz_application_1962}, Demiray \citep{demiray_note_1972}, Gent \citep{gent_forms_1958}, Holzapfel \citep{cowin_new_2004}, Mooney-Rivlin \citep{rivlin_large_1950}, Neo-Hooke \citep{treloar_stress-strain_1944}, and Ogden \citep{ogden_large_1972} models for generating the database.
By considering only physically admissible models and parameters during the offline stage, \textbf{the online pattern recognition algorithm is guaranteed to always discover models that satisfy physical constraints}, such as, for example, \textbf{stress-free undeformed configurations, growth conditions, material symmetry and polyconvexity} \citep{linden_neural_2023}. 
In this work, we impose objectivity, stress-free undeformed configurations, and isotropy on the material models, and note that polyconvexity could readily be ensured, for instance, by considering features that depend on polyconvex invariants \citep{hartmann_polyconvexity_2003,schroder_invariant_2003,linden_neural_2023}. 
A subset of the fingerprints present in the database is illustrated in \cref{fig:pattern_recognition_matrices}.
For a detailed description of the models and the corresponding parameter ranges, we refer to \cref{sec:database_supervised}.

We notice that, for fixed prescribed deformations, the fingerprint of a material depends on the material parameters $\bff(\bftheta, \bfalpha)$, and that this relationship fulfills the homogeneity property $\bff(a \bftheta, \bfalpha) = a \bff(\bftheta, \bfalpha), \ \forall a \in \Rset$.
Thus, after constructing the database of material fingerprints, the homogeneity property can be used to normalize all fingerprint vectors.
Specifically, we compute the normalized data triplets
\be
(\bar\bff^{(i)},\bar\bftheta^{(i)},\bfalpha^{(i)}) \quad \text{with} \quad \bar\bff^{(i)} = \frac{\bff^{(i)}}{\|\bff^{(i)}\|}, \ \bar\bftheta^{(i)} = \frac{\bftheta^{(i)}}{\|\bff^{(i)}\|},
\ee
where we note that $\bff(\bftheta^{(i)}/\|\bff^{(i)}\|, \bfalpha^{(i)}) = \bff(\bftheta^{(i)}, \bfalpha^{(i)})/\|\bff^{(i)}\| = \bff^{(i)}/\|\bff^{(i)}\|$.
Owing to the homogeneity property, the process of database normalization does not result in any loss of information.
The database normalization is an important feature of \MF{}. Materials that can only be distinguished by a scalar factor in $\bftheta$ are assigned to the same normalized fingerprint. In this way, the database can store information about a higher number of different materials. As we show below, the normalization factor is easily and efficiently computed in the online phase.

\begin{figure}
    \centering
    \includegraphics[width=\textwidth]{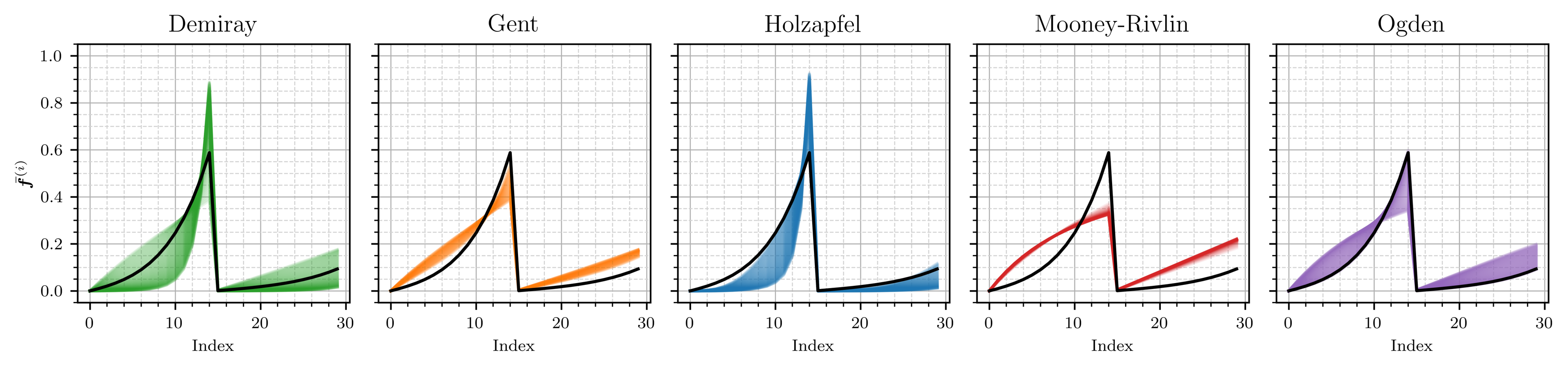}
    \caption{Illustration of supervised \MF{}. The measurement, illustrated by a black curve, is compared to all entries in the database, illustrated by colored regions. The database entries are grouped and visualized separately for each model, excluding simple models that have a single parameter only.}
    \label{fig:pattern_recognition_curves}
\end{figure}

\subsubsection{Pattern recognition algorithm}
\textbf{After constructing the database once, the offline phase is completed, and the database is used in the online phase to rapidly discover material models} for given experimental measurements.
For a material with unknown mechanical behavior, we conduct experiments considering the same prescribed deformations as those assumed during the database generation.
During the experiment, the material's fingerprint $\bff^*$ is measured.
Afterwards, a simple pattern recognition algorithm is used to identify the fingerprint in the database that is closest to the measured fingerprint.
To this end, we normalize the measured fingerprint 
$\bar\bff^* = {\bff^*}/{\|\bff^*\|}$.
To identify the closest fingerprint in the database, we compute the inner products of all database fingerprints with the measured fingerprint and select the database entry with the highest inner product
\be
\label{eq:pattern_recognition_supervised}
i^* = \argmax_{i=1, \dots ,n_d} \ \bar\bff^{(i)} \cdot \bar\bff^*.
\ee
The pattern recognition is illustrated in \cref{fig:pattern_recognition_curves}, which shows a measurement in comparison with the fingerprints in the database.
We note that, because the fingerprints are normalized, the pattern recognition algorithm can be interpreted as finding the fingerprint in the database with the smallest angle to the measured fingerprint, as illustrated in \cref{fig:abstract}.
Using the inner product as a measure of similarity is also called cosine similarity in the machine learning community \citep{bishop_pattern_2006,xia_learning_2015}.
\CHANGE{We choose this measure because of its computational efficiency.}
\CHANGE{For normalized vectors, it is $\|\bar\bff^{(i)} - \bar\bff^*\|^2 = -2  \ \bar\bff^{(i)} \cdot \bar\bff^*$. Thus, finding the maximum inner product is equivalent to finding the minimum distance between the fingerprints in the Euclidean norm, while computing the inner products requires less operations.}
The computations of the $n_d$ inner products can be interpreted as one matrix vector multiplication with $n_d$ rows and $n_f$ columns, an operation that can easily be parallelized.
Finding the maximum over the indices $i$ is not computationally demanding and can be orders of magnitude faster than solving the potentially non-convex optimization problems associated with traditional model calibration methods or the training processes used in machine learning approaches for learning mechanical material behavior \citep{fuhg_review_2024}.
Moreover, while identifying the global optimum in a non-convex optimization landscape is generally challenging, the pattern recognition procedure in \cref{eq:pattern_recognition_supervised} selects the fingerprint with the highest agreement from the database, effectively yielding the discrete global optimum within the database.
\textbf{\MF{} therefore offers a powerful alternative in scenarios where optimization-based methods struggle due to non-convexity.}

After having identified $i^*$, we use the corresponding material parameters in the database, $\bar\bftheta^{(i^*)}$ and $\bfalpha^{(i^*)}$, to determine the parameters of the tested material.
The normalization factor needed to compute the homogeneity parameters is equal to the norm of the measured fingerprint.
Thus, the identified parameters are
\be
\label{eq:rescaling_supervised}
\bftheta^* = \|\bff^*\| \bar\bftheta^{(i^*)} \quad \text{and} \quad \bfalpha^* = \bfalpha^{(i^*)}.
\ee
Here, we note that the pattern recognition process not only calibrates parameters but also identifies the most suitable model from the database to describe the data. In other words, it goes beyond fitting parameters within a predefined strain energy density function, as it also \textbf{selects the optimal functional form for representing the material behavior}.

\subsubsection{Data interpolation}
\CHANGETWO{
In this work, we focus on an entirely numerical investigation of \MF{}, assuming that the same stretches used during database generation can likewise be applied to the material in experimental settings.
This assumption may not hold in general, for example, if the material fails before the maximum stretch is reached.
However, this issue can be addressed by loading only the relevant stretch ranges from the database and interpolating the measurements to match the corresponding stretches and strains \citep{martonova_material_2025}.
}

\subsubsection{Sparsity}
Because only material models with a small number of parameters are considered during database generation, the pattern recognition algorithm consistently identifies short and interpretable mathematical expressions for the material behavior. This stands in contrast to black-box machine learning approaches. We note that, when the database contains models with varying numbers of parameters, a sparsity-promoting regularization term \citep{frank_statistical_1993,tibshirani_regression_1996,mcculloch_sparse_2024,flaschel_non-smooth_2025} may be incorporated into the pattern recognition algorithm. For example, one may introduce
\be
i^* = \argmax_{i=1, \dots ,n_d} \ \bar\bff^{(i)} \cdot \bar\bff^* - \CHANGE{\xi} \ \|\bftheta^{(i)}\|_0,
\ee
with $\CHANGE{\xi} > 0$ and $\|\bftheta^{(i)}\|_0$ denoting the number of nonzero parameters in $\bftheta^{(i)}$. In this way, models with higher sparsity are favored over more complex alternatives during pattern recognition. In this work, however, all models in the database are sparse, and we therefore use \cref{eq:pattern_recognition_supervised} without additional regularization.

\subsubsection{Database compression and accelerated pattern recognition}

\CHANGETWO{\MF{} may encounter limitations in applications requiring models with a very large number of parameters, where sampling the corresponding parameter space becomes computationally intractable.}
While we do not explore this possibility in the present work, we note that future efforts may employ database compression techniques \citep{mcgivney_svd_2014} to reduce the number of fingerprints, for example, by removing redundant or less significant entries. Alternatively, dimensionality reduction methods could be applied to directly reduce the dimensionality of the fingerprints themselves. Additionally, approaches to accelerate the pattern recognition algorithm may be investigated, for instance, by partitioning the database into subgroups with representative fingerprints. Initially, the most relevant subgroup would be identified, followed by a more refined search within that subgroup.

\subsection{Unsupervised \MF{}}

In the experiments discussed previously, the deformation fields are homogeneous across the specimens.
As a result, these experiments provide limited insight into how the material behaves under varied deformation states.
\textbf{A material model discovered from uniaxial tension and simple shear tests, for instance, may not accurately predict the material's behavior under other deformation conditions.}
Recognizing this limitation, experiments featuring heterogeneous deformation fields are increasingly popular for calibrating \citep{grediac_principle_1989,avril_overview_2008,pierron_towards_2020}, discovering \citep{flaschel_unsupervised_2021}, and learning \citep{thakolkaran_nn-euclid_2022,benady_unsupervised_2024,wiesheier_versatile_2024} material models, marking a paradigm shift toward what is now referred to as Material \mbox{Testing 2.0} \citep{pierron_towards_2020}.

In this work, we extend these concepts to \MF{}.
We first design a standardized experiment using a complexly shaped specimen.
Specifically, we focus on a plate with a hole subjected to biaxial tension, although other, optimized geometries and loading conditions are also possible \citep{bensoe_optimal_1989,grediac_t-shaped_1998,souto_numerical_2016,bertin_optimization_2016,chamoin_coupling_2020,ghouli_topology_2025}.
We then construct a material fingerprint database by simulating these experiments across various material models and parameters.
Given that the experimental testing of complexly shaped specimens does not yield labeled strain-stress data pairs, we cannot design the fingerprints in the same way as in supervised \MF{}.
While the fingerprints in the supervised context consist of stress measurements for different deformation states, we select net reaction force measurements at the specimen's boundary and displacement measurements on its surface as the material fingerprints in the unsupervised context.
Once the fingerprint database is generated and normalized in the offline phase, a pattern recognition algorithm can be employed to identify material models for new, unseen materials.

\CHANGE{
Existing methods for unsupervised model calibration and discovery can be broadly divided into two categories: methods that minimize the discrepancy between measured and simulated displacements, such as Finite Element Model Updating, and methods that minimize the residuals of the weak formulation of the linear momentum balance, such as the Virtual Fields Method and Equilibrium Gap Method \citep{avril_overview_2008}. Displacement-based methods are generally robust, as the displacement fields do not need to be spatially differentiated \citep{romer_reduced_2024}. Weak formulation-based methods, on the other hand, are typically computationally efficient because they avoid repeatedly solving the boundary value problem.
As \MF{} is displacement-based, it combines the robustness of the former with the efficiency of the latter.
}

In the following, we provide a detailed description of the experimental setup, the characteristics of the fingerprints, the material models assumed during database generation, and the pattern recognition algorithm employed in this study.

\subsubsection{Experiment design}

\begin{figure}[h!]
\centering
\includegraphics[width=0.7\linewidth]{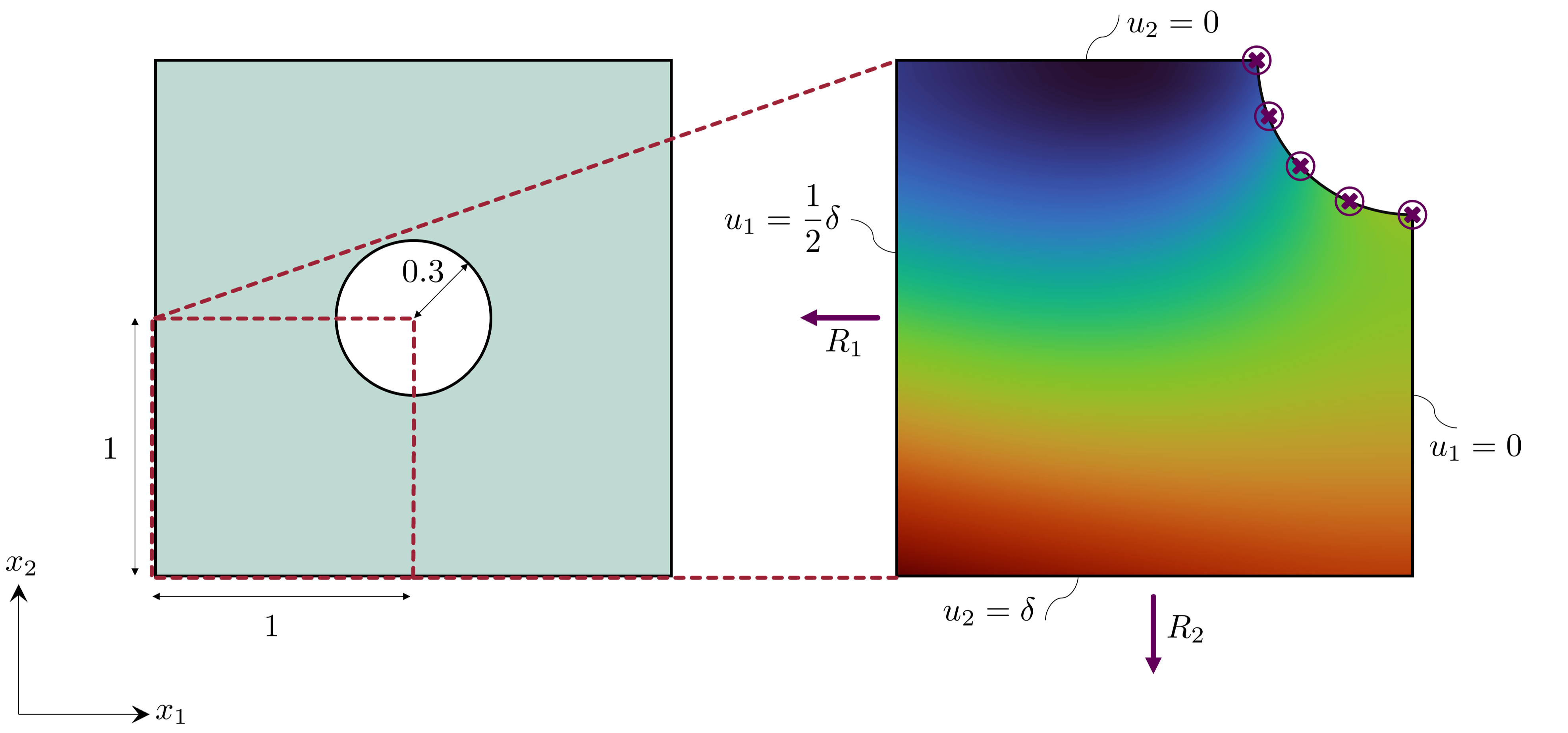}
\caption{Specimen geometry and loading conditions for unsupervised \MF{}. During the experiment, the reaction forces $R_1$ and $R_2$, along with displacements at selected points near the central hole, are measured for all load steps.}
\label{fig:plate}
\end{figure}

In this study, we consider a plate with a hole under biaxial loading and plane strain conditions, see \cref{fig:plate}.
To generate fingerprints artificially, we simulate the deformation of the plate using the Finite Element Method in displacement control.
Due to the symmetry conditions, it is sufficient to consider only one quarter of the specimen, for which we fix the vertical displacement of the top boundary and the horizontal displacement of the right boundary.
Further, we prescribe a vertical displacement of $\delta$ to the bottom boundary and a horizontal displacement of $\frac{1}{2} \delta$ to the left boundary as Dirichlet boundary conditions, and we assume homogeneous Neumann boundary conditions for all other degrees of freedom at the boundary.
We increase $\delta$ during $n_t = 10$ equidistant load steps until reaching a maximum of $\delta=0.3$.

For fingerprint generation, we assume unit dimensions for the plate.
However, this choice does not impose a restriction on the actual experimental setup.
If the real specimen tested during the online stage has the same geometry but is scaled by a constant factor, the same fingerprint database remains valid, provided the experimental data \CHANGE{are} scaled accordingly before applying the pattern recognition algorithm.

\subsubsection{Fingerprint definition}

\CHANGE{Depending on} the mechanical behavior of the material, the deformation of the plate yields different displacement fields over the specimen domain and different net reaction forces at the specimen boundaries.
Thus, we define the materials' fingerprints as a combination of displacement and net reaction force measurements.
Specifically, we define the vector $\bff_R \in \Rset^{2 \cdot n_t}$, containing the reaction forces $R_1$ and $R_2$ for each of \CHANGE{the} $n_t$ load steps, see \cref{fig:plate}.
In addition, we define $\bff_u \in \Rset^{2 \cdot n_u \cdot n_t}$, a vector containing $n_u$ displacement measurements for both $x_1$- and $x_2$-directions at different points at the hole for each load step.
\CHANGE{Displacement measurement can, for example, be acquired through digital image correlation. However, because digital image correlation measurements near specimen boundaries can be erroneous in real experiments, displacement measurements may instead be acquired using point-tracking technologies. Alternatively, measurements close to the hole, but not exactly at the boundary, may be considered when defining the fingerprints.}

The resulting forces and displacements depend on the material parameters.
\CHANGE{Under pure displacement control, scaling the strain energy density by a factor results in a proportional scaling of the reaction forces, while the displacement field remains unchanged. Thus,} forces fulfill the homogeneity property $\bff^{(i)}_R(a \bftheta, \bfalpha) = a \bff^{(i)}_R(\bftheta, \bfalpha), \ \forall a \in \Rset$, and the displacements are invariant to a scaling of the homogeneity parameters $\bff^{(i)}_u(a \bftheta, \bfalpha) = \bff^{(i)}_u(\bftheta, \bfalpha), \ \forall a \in \Rset$.
The material fingerprints $\bff$ are ultimately defined as the concatenation of the vectors $\bff_R$ and $\bff_u$.
Since the material exhibits a range of different deformation states within a single experiment, \textbf{the fingerprints obtained in the unsupervised \MF{} encode more comprehensive information about the mechanical characteristics of the material compared to the supervised approach}.
This richness not only aids in the discovery of material models from a single experiment but also allows for model predictions beyond the confines of uniaxial tension and simple shear states.

\subsubsection{Fingerprint database generation}
To demonstrate the functionality of \MF{} in the unsupervised setting, we focus on isotropic and compressible hyperelasticity in this work.
We generate a database of fingerprints by considering compressible versions of the Blatz-Ko, Demiray, Gent, Holzapfel, Mooney-Rivlin, and Neo-Hooke models. A detailed description of the models and the discretization of the material parameters is provided in \cref{sec:database_unsupervised}.
For each model and each parameter realization, we simulate the deformation of the plate and compute forces $\bff^{(i)}_R$ and displacements $\bff^{(i)}_u$, where $i=1, \dots ,n_d$.
We normalize the computed vectors $\bar \bff^{(i)}_R = \bff^{(i)}_R/\|\bff^{(i)}_R\|$, $\bar \bff^{(i)}_u = \bff^{(i)}_u/\|\bff^{(i)}_u\|$ and scale the corresponding homogeneity parameters $\bar \bftheta^{(i)} = \bftheta^{(i)}/\|\bff^{(i)}_R\|$.
After normalization, we define the normalized fingerprints $\bar \bff^{(i)}$ by concatenating both the normalized reaction forces $\bar \bff^{(i)}_R$ and displacements $\bar \bff^{(i)}_u$.


\subsubsection{Pattern recognition algorithm}
After having generated the fingerprint database entries $\bar \bff^{(i)}$, they are used to discover the material model for a measured fingerprint $\bff^*$ of a previously unseen material.
The fingerprint of an unseen material can be measured by conducting an experiment with the same previously described loading conditions.
During the experiment, the reaction forces $\bff_R^*$ are measured using load cells and the displacements $\bff_u^*$ are measured using, for example, digital image correlation or point markers near the hole.
The measured fingerprint \CHANGE{is} normalized by dividing $\bff^*_R$ and $\bff^*_u$ by their norms, and vector concatenation yields the normalized measured fingerprint $\bar\bff^*$.
Then, we leverage the pattern recognition algorithm
\be
\label{eq:pattern_recognition_unsupervised}
i^* = \argmax_{i=1, \dots ,n_d} \ \bar\bff^{(i)} \cdot \bar\bff^*,
\ee
to identify $i^*$, that is, the index in the database showing the highest correlation with the measurement.
We arrive at the discovered material model by rescaling the homogeneity parameters with the norm of the force measurements
\be
\label{eq:rescaling_unsupervised}
\bftheta^* = \|\bff^*_R\| \bar\bftheta^{(i^*)} \quad \text{and} \quad \bfalpha^* = \bfalpha^{(i^*)}.
\ee

We note that the measure of similarity in \cref{eq:pattern_recognition_unsupervised} may also be written as $\bar\bff^{(i)} \cdot \bar\bff^* = \bar\bff^{(i)}_R \cdot \bar\bff^*_R + \bar\bff^{(i)}_u \cdot \bar\bff^*_u$.
It is related to the angles between $\bff^{(i)}_R$ and $\bff^*_R$ and between $\bff^{(i)}_u$ and $\bff^*_u$.
Thus, the measure does not take the absolute values of the forces and displacements into account.
While the absolute values of the forces affect the discovered model through the rescaling procedure in \cref{eq:rescaling_unsupervised}, the absolute values of the displacements are not affecting the model discovery.
In our numerical experiments, we found that the measure \cref{eq:pattern_recognition_unsupervised} is sufficient for accurately identifying a suitable fingerprint in the database.
In the future, however, other measures of similarity, such as, for example, $\bar\bff^{(i)}_R \cdot \bar\bff^*_R - \lambda_u \|\bff^{(i)}_u - \bff^*_u \|$ with $\lambda_u > 0$, may be considered to take the absolute values of the displacement measurements into account.

\section{Benchmarks}
\label{sec:benchmarks}

\subsection{Supervised \MF{}}
\label{sec:benchmarks_supervised}

\CHANGETWO{To demonstrate the numerical functionality of \MF{}, we focus on numerically generated data in this work.
We test the supervised \MF{} method on data genrated using five benchmark material models, see \cref{tab:benchmarks_supervised}.}
We measure the material fingerprints following the same experimental settings assumed during the database generation in the offline stage.
To emulate real experiments, we add independent, normally distributed noise with zero mean to the stress measurements before normalization.
Two noise levels are investigated by setting the standard deviation to $1 \%$ and $5 \%$ of the maximum absolute stress measurement.
The fast pattern recognition algorithm is then used, for each benchmark and each noise level, to discover the material model in the database, see \cref{sec:database_supervised}, that best matches the measured fingerprint.

To quantify the mismatch between a true and discovered strain energy density, we introduce the error
\be
\label{eq:error_incompressible}
E_{\text{incompr}} = \frac{
\int_a^b \int_a^b \left| W_{\text{true}}(\lambda_1,\lambda_2,\lambda_3) - W_{\text{disc}}(\lambda_1,\lambda_2,\lambda_3) \right| \ \mathrm{d} \lambda_1 \mathrm{d} \lambda_2
}{
\int_a^b \int_a^b \left| W_{\text{true}}(\lambda_1,\lambda_2,\lambda_3) \right| \ \mathrm{d} \lambda_1 \mathrm{d} \lambda_2
}
\quad \text{s.t.} \quad \lambda_3 = 1/[\lambda_1 \lambda_2]
,
\ee
where $W_{\text{true}}$ and $W_{\text{disc}}$ are the true and discovered strain energy densities, respectively, expressed as functions of the principal stretches.
The advantage of the error definition in \cref{eq:error_incompressible} over other measures, such as the mean squared error or the coefficient of determination $R^2$, is that $E_{\text{incompr}}$ does not depend on the nature of the experiment.
That is, when computing $E_{\text{incompr}}$, the strain energy densities are evaluated not only for the deformation states present in the experimental dataset.
Instead, $E_{\text{incompr}}$ quantifies the error even for unseen deformations beyond the scope of uniaxial tension and simple shear, making it a more general measure applicable across different experimental conditions.
Here, we heuristically choose $a=0.75$ and $b=1.25$.
We note, though, that $W_{\text{true}}$ is only known if artificially generated data are available. For experimental data, the mean squared error or the coefficient of determination $R^2$ would be the appropriate mismatch measures.
\CHANGETWO{Thus, in addition to the error measure above, we compute the coefficient of determination as
\begin{equation}
\label{eq:r2}
R^2 = 1 - \frac{\sum_{i=1}^{n_f} \left( f^*_{i} - f_{i} \right)^2}{\sum_{i=1}^{n_f} \left( f^*_{i} - \text{mean}(\bff^*) \right)^2}.
\end{equation}
Here, $\bff^*$ denotes the experimentally measured fingerprint vector without normalization, that is, the stress values of all considered experiments for all considered stretch and strain levels,
$\text{mean}(\bff^*)$ is the mean value of the experimental stresses and
$\bff$ is the fingerprint of the model, that is, the stress values predicted by the model.}

\begin{table}[h!]
\caption{Strain energy density functions discovered through supervised \MF{}. \CHANGE{Different colors indicate different material models in the database.}}
\label{tab:benchmarks_supervised}
\centering
\begin{tabular}{|ll|l|r|r|}
\hline
\multicolumn{2}{|l|}{Benchmarks\vphantom{$\int_a^b$}} & \multicolumn{1}{c|}{Strain energy density $W$} & \multicolumn{1}{c|}{$E_{\text{incompr}}$} & \multicolumn{1}{c|}{\CHANGETWO{$R^2$}} \\ \hline

\rowcolor{BlatzKo!12}
\multicolumn{1}{|l|}{Blatz-Ko\vphantom{$\int_a^b$}} & Truth & $50.00 [I_2 - 3] - p [J-1]$ & -- & -- \\ \cline{2-5}
\rowcolor{BlatzKo!12}
\multicolumn{1}{|l|}{\vphantom{$\int_a^b$}} & $0\%$ Noise &  $50.00 [I_2 - 3] - p [J-1]$& \num{0.00} & 1.0000 \\ \cline{2-5}
\rowcolor{BlatzKo!12}
\multicolumn{1}{|l|}{\vphantom{$\int_a^b$}} & $1\%$ Noise & $50.35 [I_2 - 3] - p [J-1]$ & \num{6.98e-03} & 0.9998 \\ \cline{2-5}
\rowcolor{BlatzKo!12}
\multicolumn{1}{|l|}{\vphantom{$\int_a^b$}} & $5\%$ Noise & $49.24 [I_2 - 3] - p [J-1]$ & \num{1.53e-02}  & 0.9991 \\ \hline

\rowcolor{Demiray!12}
\multicolumn{1}{|l|}{Demiray\vphantom{$\int_a^b$}} & Truth & $10.00 \left[\exp(8.00 [I_1 - 3]) - 1\right] - p [J-1]$ & -- & -- \\ \cline{2-5}
\rowcolor{Demiray!12}
\multicolumn{1}{|l|}{\vphantom{$\int_a^b$}} & $0\%$ Noise & $10.00 \left[\exp(8.00 [I_1 - 3]) - 1\right] - p [J-1]$ & \num{0.00} & 1.0000 \\ \cline{2-5}
\rowcolor{Demiray!12}
\multicolumn{1}{|l|}{\vphantom{$\int_a^b$}} & $1\%$ Noise & $10.02 \left[\exp(8.00 [I_1 - 3]) - 1\right] - p [J-1]$ & \num{2.11e-03} & 1.0000 \\ \cline{2-5}
\rowcolor{Demiray!12}
\multicolumn{1}{|l|}{\vphantom{$\int_a^b$}} & $5\%$ Noise &  $19.83 \left[\exp(7.00 [I_1 - 3]) - 1\right] - p [J-1]$ & \num{2.86e-01} & 0.9958 \\ \hline

\rowcolor{MooneyRivlin!12}
\multicolumn{1}{|l|}{Mooney-Rivlin\vphantom{$\int_a^b$}} & Truth & $10.00 [I_1 - 3] + 40.00 [I_2 - 3] - p [J-1]$ & -- & -- \\ \cline{2-5}
\rowcolor{MooneyRivlin!12}
\multicolumn{1}{|l|}{\vphantom{$\int_a^b$}} & $0\%$ Noise & $10.00 [I_1 - 3] + 40.00 [I_2 - 3] - p [J-1]$ & \num{0.00} & 1.0000 \\ \cline{2-5}
\rowcolor{MooneyRivlin!12}
\multicolumn{1}{|l|}{\vphantom{$\int_a^b$}} & $1\%$ Noise & $9.47 [I_1 - 3] + 40.73 [I_2 - 3] - p [J-1]$ & \num{3.70e-03} & 1.0000 \\ \cline{2-5}
\rowcolor{MooneyRivlin!12}
\multicolumn{1}{|l|}{\vphantom{$\int_a^b$}} & $5\%$ Noise & $12.28 [I_1 - 3] + 36.83 [I_2 - 3] - p [J-1]$ & \num{1.69e-02} & 0.9996 \\ \hline

\rowcolor{NeoHooke!12}
\multicolumn{1}{|l|}{Neo-Hooke\vphantom{$\int_a^b$}} & Truth & $10.00 [I_1 - 3] - p [J-1]$ & -- & -- \\ \cline{2-5}
\rowcolor{NeoHooke!12}
\multicolumn{1}{|l|}{\vphantom{$\int_a^b$}} & $0\%$ Noise & $10.00 [I_1 - 3] - p [J-1]$ & \num{0.00} & 1.0000 \\ \cline{2-5}
\rowcolor{NeoHooke!12}
\multicolumn{1}{|l|}{\vphantom{$\int_a^b$}} & $1\%$ Noise & $9.97 [I_1 - 3] - p [J-1]$ & \num{3.14e-03} & 1.0000 \\ \cline{2-5}
\rowcolor{Ogden!12}
\multicolumn{1}{|l|}{\vphantom{$\int_a^b$}} & $5\%$ Noise & $12.50 [\lambda_1^{1.80} + \lambda_2^{1.80} + \lambda_3^{1.80} - 3] - p [J-1]$ & \num{1.11e-02} & 0.9996 \\ \hline

\rowcolor{Ogden!12}
\multicolumn{1}{|l|}{Ogden\vphantom{$\int_a^b$}} & Truth & $5.00[\lambda_1^{8.00} + \lambda_2^{8.00} + \lambda_3^{8.00} - 3] - p [J-1]$ & -- & -- \\ \cline{2-5}
\rowcolor{Ogden!12}
\multicolumn{1}{|l|}{\vphantom{$\int_a^b$}} & $0\%$ Noise & $5.00 [\lambda_1^{8.00} + \lambda_2^{8.00} + \lambda_3^{8.00} - 3] - p [J-1]$& \num{0.00} & 1.0000 \\ \cline{2-5}
\rowcolor{Ogden!12}
\multicolumn{1}{|l|}{\vphantom{$\int_a^b$}} & $1\%$ Noise & $4.76 [\lambda_1^{8.10} + \lambda_2^{8.10} + \lambda_3^{8.10} - 3] - p [J-1]$& \num{1.24e-02} & 0.9999 \\ \cline{2-5}
\rowcolor{Ogden!12}
\multicolumn{1}{|l|}{\vphantom{$\int_a^b$}} & $5\%$ Noise & $5.22 [\lambda_1^{8.00} + \lambda_2^{8.00} + \lambda_3^{8.00} - 3] - p [J-1]$ & \num{4.49e-02} & 0.9965 \\ \hline

\end{tabular}
\end{table}

\subsection{Unsupervised \MF{}}
We test the unsupervised \MF{} method on artificially generated data using four benchmark material models, see \cref{tab:benchmarks_unsupervised}.
Subsequently, we proceed as described in the previous \cref{sec:benchmarks_supervised}.
The material model database for the unsupervised case is summarized in \cref{sec:database_unsupervised}. 
In addition to the quantification of the mismatch between the true and discovered strain energy density in the incompressible case, we introduce an error measure for the compressible case as 
\be
E_{\text{compr}} = \frac{
\int_a^b \int_a^b \int_a^b \left| W_{\text{true}}(\lambda_1,\lambda_2,\lambda_3) - W_{\text{disc}}(\lambda_1,\lambda_2,\lambda_3) \right| \ \mathrm{d} \lambda_1 \mathrm{d} \lambda_2 \mathrm{d} \lambda_3
}{
\int_a^b \int_a^b \int_a^b \left| W_{\text{true}}(\lambda_1,\lambda_2,\lambda_3) \right| \ \mathrm{d} \lambda_1 \mathrm{d} \lambda_2 \mathrm{d} \lambda_3
},
\ee
where the stretches $\lambda_1$, $\lambda_2$ and $\lambda_3$ can have arbitrary values greater than zero.
\CHANGETWO{Additionally, we define the coefficient of determination analogously to \cref{eq:r2}, where $\bff^*=[\bff^*_R; \bff^*_u]$ is the measured fingerprint vector containing the reaction force and displacement measurements without normalization and $\bff=[\bff_R; \bff_u]$ is the fingerprint predicted by the model.}

\begin{table}[h!]
\caption{Strain energy density functions discovered through unsupervised \MF{}. \CHANGE{Different colors indicate different material models in the database.}}
\label{tab:benchmarks_unsupervised}
\centering
\resizebox{16cm}{!}{
\begin{tabular}{|ll|l|r|r|r|}
\hline
\multicolumn{2}{|l|}{Benchmarks\vphantom{$\int_a^b$}} & \multicolumn{1}{c|}{Strain energy density $W$} & \multicolumn{1}{c|}{$E_{\text{compr}}$} & \multicolumn{1}{c|}{$E_{\text{incompr}}$} & \multicolumn{1}{c|}{\CHANGETWO{$R^2$}} \\ \hline

\rowcolor{BlatzKo!12}
\multicolumn{1}{|l|}{Blatz-Ko\vphantom{$\int_a^b$}} & Truth & $50.00 [\bar I_2 - 3] + 5.00 [J - 1]^2$ & -- & -- & -- \\ \cline{2-6}
\rowcolor{BlatzKo!12}
\multicolumn{1}{|l|}{\vphantom{$\int_a^b$}} & $0\%$ Noise & $50.00 [\bar I_2 - 3] + 5.00 [J - 1]^2$ & $0.00$ & $0.00$ & 1.0000 \\ \cline{2-6}
\rowcolor{BlatzKo!12}
\multicolumn{1}{|l|}{\vphantom{$\int_a^b$}} & $1\%$ Noise & $49.52 [\bar I_2 - 3] + 5.05 [J - 1]^2$ & $8.61 \cdot 10^{-3}$ & $9.63 \cdot 10^{-3}$ & 0.9994 \\ \cline{2-6}
\rowcolor{BlatzKo!12}
\multicolumn{1}{|l|}{\vphantom{$\int_a^b$}} & $5\%$ Noise & $48.25 [\bar I_2 - 3] + 5.19 [J - 1]^2$ & $3.13 \cdot 10^{-2}$ & $3.50 \cdot 10^{-2}$ & 0.9845 \\ \hline

\rowcolor{Demiray!12}
\multicolumn{1}{|l|}{Demiray\vphantom{$\int_a^b$}} & Truth & $10.00 [\exp(8.00 [\bar I_1 - 3]) - 1] + 5.00 [J - 1]^2$ & -- & -- & -- \\ \cline{2-6}
\rowcolor{Demiray!12}
\multicolumn{1}{|l|}{\vphantom{$\int_a^b$}} & $0\%$ Noise & $10.00 [\exp(8.00 [\bar I_1 - 3]) - 1] + 5.00 [J - 1]^2$ & $0.00$ & $0.00$ & 1.0000 \\ \cline{2-6}
\rowcolor{Demiray!12}
\multicolumn{1}{|l|}{\vphantom{$\int_a^b$}} & $1\%$ Noise & $10.01 [\exp(8.00 [\bar I_1 - 3]) - 1] + 5.00 [J - 1]^2$ & $6.27 \cdot 10^{-4}$ & $6.27 \cdot 10^{-4}$ & 0.9995 \\ \cline{2-6}
\rowcolor{Demiray!12}
\multicolumn{1}{|l|}{\vphantom{$\int_a^b$}} & $5\%$ Noise & $9.79 [\exp(8.10 [\bar I_1 - 3]) - 1] + 5.44 [J - 1]^2$ & $4.65 \cdot 10^{-3}$ & $7.74 \cdot 10^{-2}$ & 0.9879 \\ \hline

\rowcolor{MooneyRivlin!12}
\multicolumn{1}{|l|}{Mooney-Rivlin\vphantom{$\int_a^b$}} & Truth & $10.00 [\bar I_1 - 3] + 40.00 [\bar I_2 - 3] + 20.00 [J - 1]^2$ & -- & -- & -- \\ \cline{2-6}
\rowcolor{MooneyRivlin!12}
\multicolumn{1}{|l|}{\vphantom{$\int_a^b$}} & $0\%$ Noise & $10.00 [\bar I_1 - 3] + 40.00 [\bar I_2 - 3] + 20.00 [J - 1]^2$ & $0.00$ & $0.00$ & 1.0000 \\ \cline{2-6}
\rowcolor{MooneyRivlin!12}
\multicolumn{1}{|l|}{\vphantom{$\int_a^b$}} & $1\%$ Noise & $14.28 [\bar I_1 - 3] + 34.69 [\bar I_2 - 3] + 20.40 [J - 1]^2$ & $1.63 \cdot 10^{-2}$ & $2.12 \cdot 10^{-2}$ & 0.9995 \\ \cline{2-6}
\rowcolor{MooneyRivlin!12}
\multicolumn{1}{|l|}{\vphantom{$\int_a^b$}} & $5\%$ Noise & $29.02 [\bar I_1 - 3] + 15.63 [\bar I_2 - 3] + 22.32 [J - 1]^2$ & $8.43 \cdot 10^{-2}$ & $1.05 \cdot 10^{-1}$ & 0.9872 \\ \hline

\rowcolor{NeoHooke!12}
\multicolumn{1}{|l|}{Neo-Hooke\vphantom{$\int_a^b$}} & Truth & $10.00 [\bar I_1 - 3] + 20.00 [J - 1]^2$ & -- & -- & -- \\ \cline{2-6}
\rowcolor{NeoHooke!12}
\multicolumn{1}{|l|}{\vphantom{$\int_a^b$}} & $0\%$ Noise & $10.00 [\bar I_1 - 3] + 20.00 [J - 1]^2$ & $0.00$ & $0.00$ & 1.0000 \\ \cline{2-6}
\rowcolor{NeoHooke!12}
\multicolumn{1}{|l|}{\vphantom{$\int_a^b$}} & $1\%$ Noise & $9.99 [\bar I_1 - 3] + 19.98 [J - 1]^2$ & $9.07 \cdot 10^{-4}$ & $9.07 \cdot 10^{-4}$ & 0.9996 \\ \cline{2-6}
\rowcolor{Demiray!12}
\multicolumn{1}{|l|}{\vphantom{$\int_a^b$}} & $5\%$ Noise & $111.18 \left[\exp(0.10 [\bar I_1 - 3]) - 1\right] + 17.93 [J - 1]^2$ & $7.25 \cdot 10^{-2}$ & $1.31 \cdot 10^{-1}$ & 0.9911 \\ \hline

\end{tabular}
}
\end{table}

\section{Results}
For both the supervised and unsupervised \MF{}, we observe consistent trends across all benchmark material models tested under varying noise levels, 0\%, 1\%, and 5\%, as shown in \cref{tab:benchmarks_supervised} and \cref{tab:benchmarks_unsupervised}. Notably, at 0\% noise, the discovered strain energy densities perfectly match the true models, yielding errors that vanish up to machine precision for each case. As expected, the discovered parameters progressively deviate from the true parameters with increasing noise levels. At 1\% noise, the errors are generally small, remaining around \(10^{-3}\), which illustrates the method's robustness against minor disturbances. At 5\% noise, more substantial deviations are observed, which is expected given that the data are significantly noisy. \CHANGETWO{The coefficient of determination $R^2$, computed on the training data, exceeds 0.98 in all examples.}

\begin{figure}[h!]
\centering
\begin{subfigure}{0.7\textwidth}
\centering
\includegraphics[width=\linewidth]{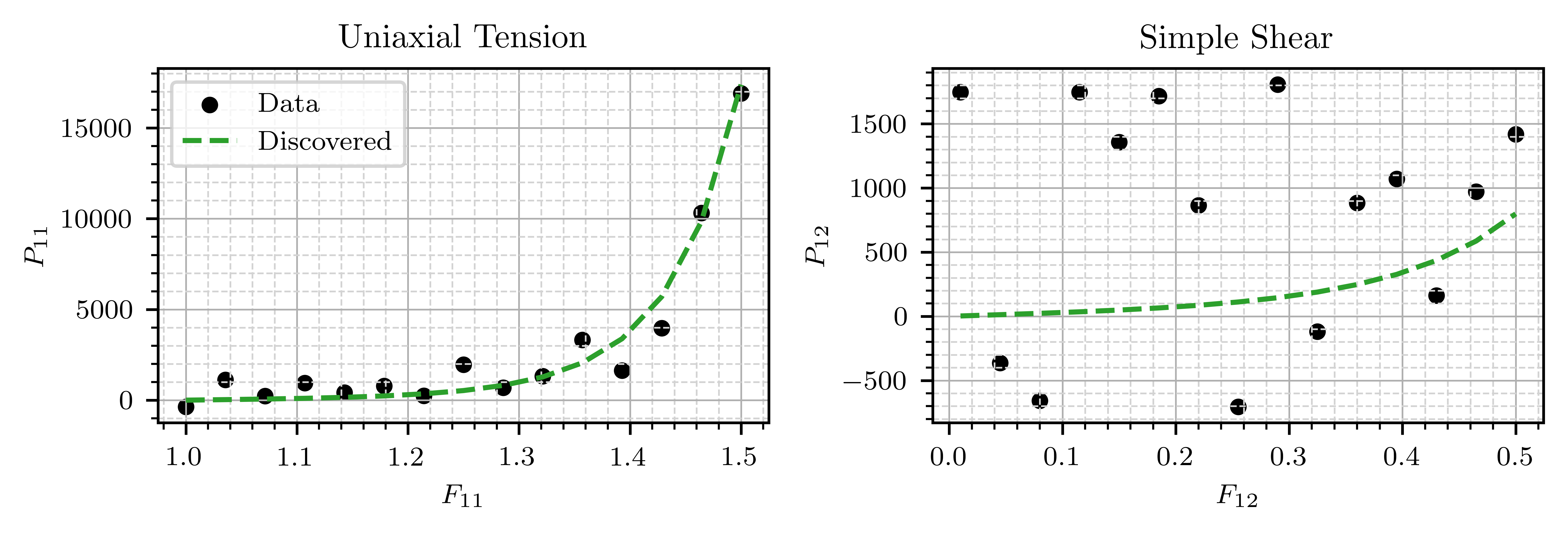}
\caption{Artificially generated data using the Demiray model with 5\% noise, and the discovered Demiray model.}
\end{subfigure}\\%
\begin{subfigure}{0.68\textwidth}
\centering
\hspace*{0.1cm}
\includegraphics[width=\linewidth]{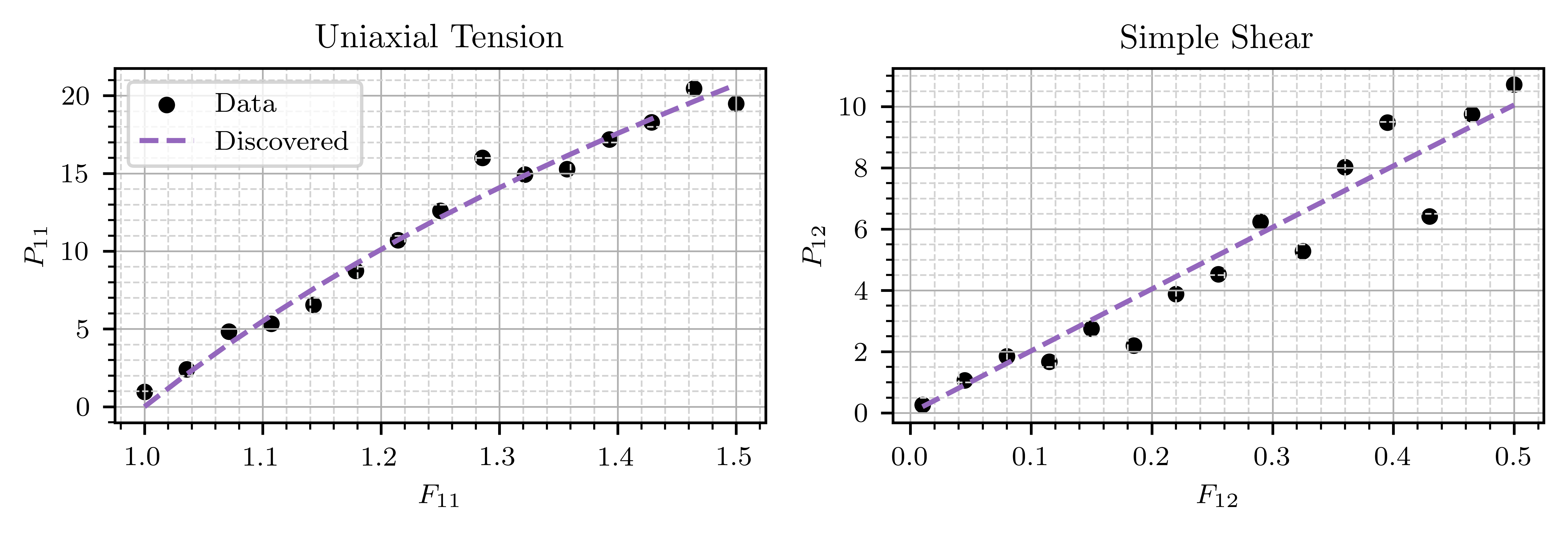}
\caption{Artificially generated data using the Neo-Hooke model with 5\% noise, and the discovered Ogden model.}
\end{subfigure}\\%
\caption{Stress response of the discovered models in comparison to the uniaxial tension (left) and simple shear (right) data with the highest noise level.
}
\label{fig:stress_strain}
\end{figure}

In the supervised setting, the largest error is observed for the Demiray model. This can be explained by the severe appearance of noise in the simple shear data for this example, see \cref{fig:stress_strain}, which shows the stress response of the discovered Demiray model in comparison to the data. A particularly interesting case in the supervised setting is the Neo-Hooke model. At 5\% noise, the discovered model transitions to an Ogden form, specifically \(12.50 [\lambda_1^{1.80} + \lambda_2^{1.80} + \lambda_3^{1.80} - 3] - p [J-1]\). The Ogden model is mathematically equivalent to the Neo-Hooke model if the exponent parameter is equal to two. Thus, the Ogden model with $\alpha_6 = 1.80$ surrogates the Neo-Hooke model, which is reflected in the good agreement between the Ogden model and the data stemming from the Neo-Hooke model in \cref{fig:stress_strain}. \CHANGETWO{As a complementary study, we show in \cref{sec:database_corrupted} the effect of excluding the ground truth models from the database, such that \MF{} cannot recover them but instead discovers alternative material models.}

\begin{figure}[h!]
\centering
\begin{subfigure}{0.6\textwidth}
\includegraphics[width=\linewidth]{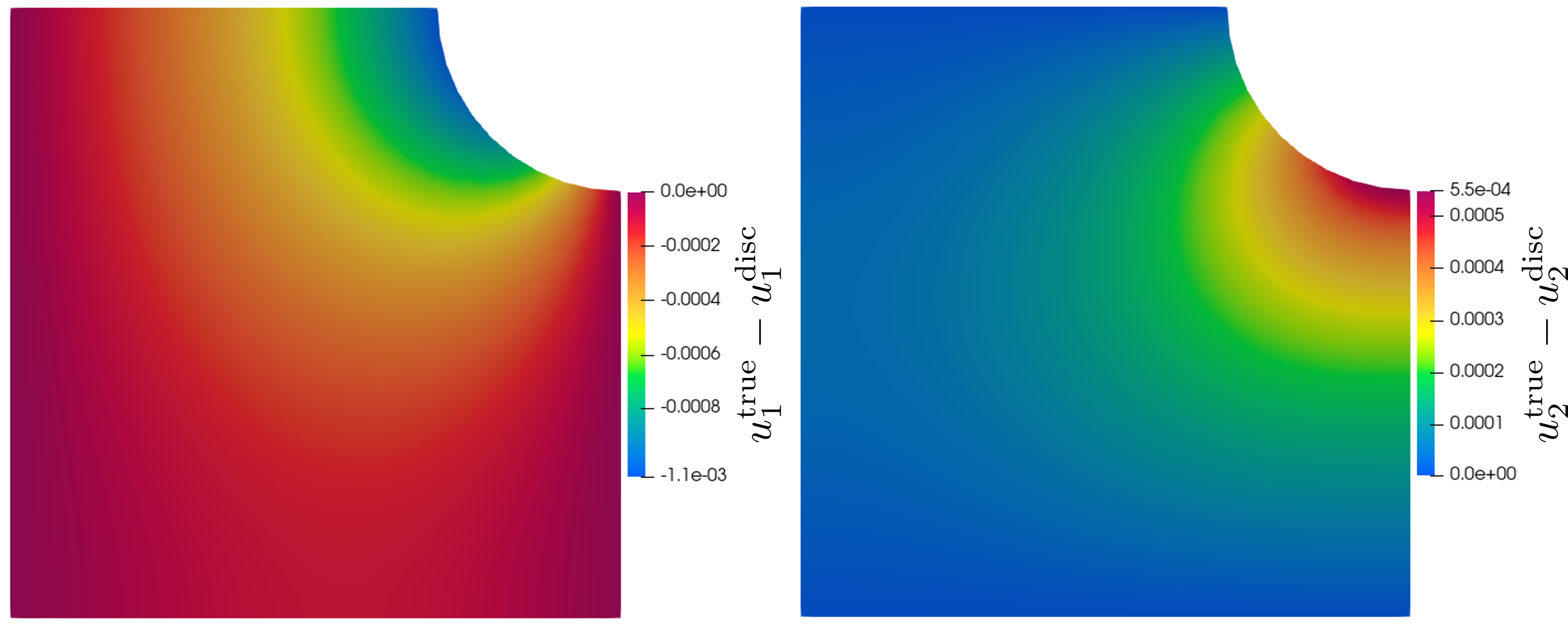}
\caption{Displacement difference in $x_1$- and $x_2$-direction.}
\end{subfigure}%
\hspace{0.2cm}
\begin{subfigure}{0.33\textwidth}
\includegraphics[width=\linewidth]{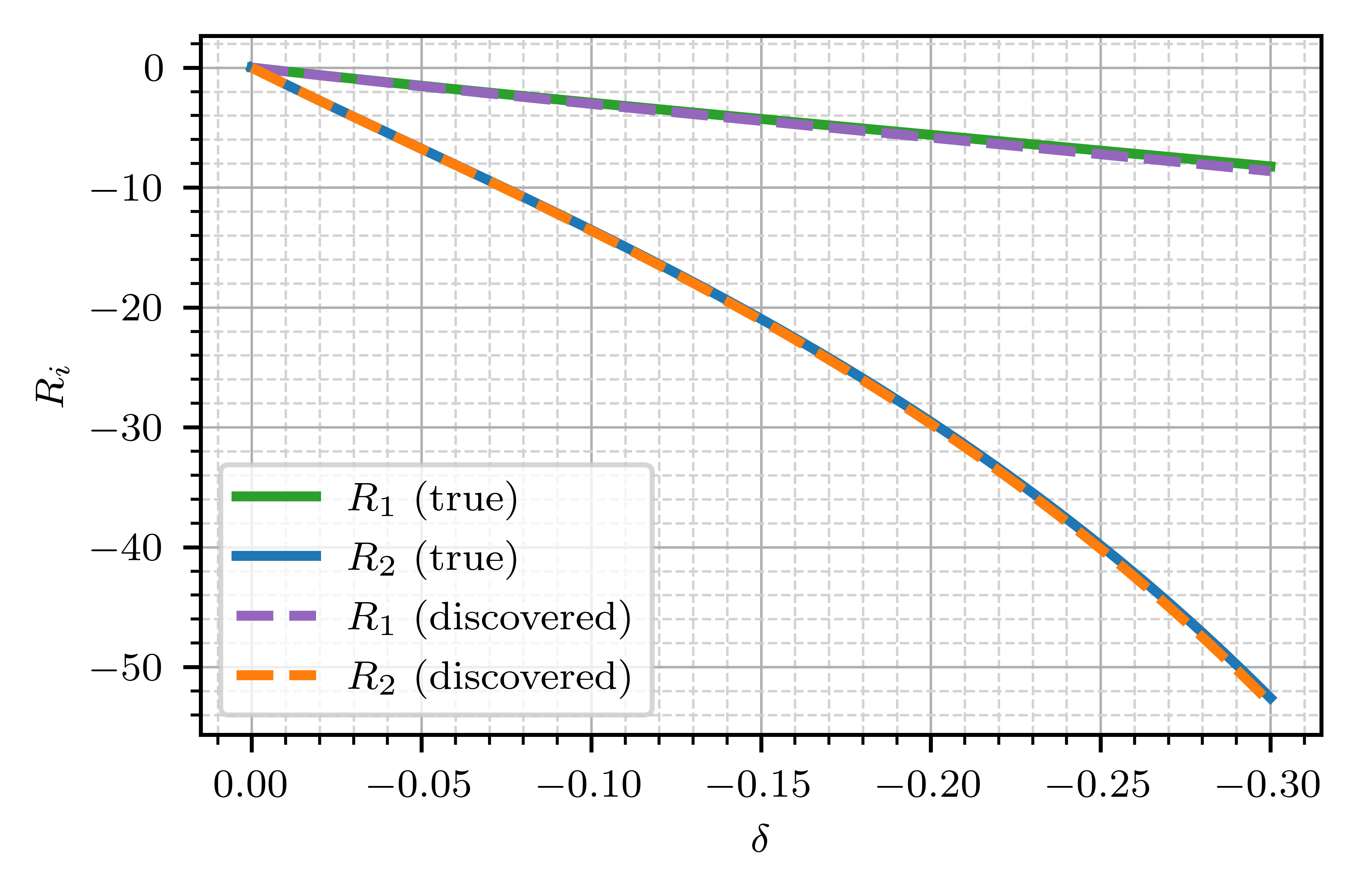}
\caption{Net reaction forces.}
\end{subfigure}%
\vspace{0.0cm}
\caption{Comparison of the simulated displacements (a) and reaction forces (b) of the true and discovered Demiray model for $5 \%$ noise.
}
\label{fig:displacement_reaction_force_Demiray}
\end{figure}

In the unsupervised \MF{}, the results follow similar patterns. Notably, \(E_{\text{incompr}}\) consistently remains greater than \(E_{\text{compr}}\) due to the incompressibility condition not being enforced during simulations. At \(5\,\%\) noise, the data produced by the Neo-Hooke model is identified as the Demiray model, demonstrating the method's ability to adapt and discover alternative representations that can still achieve good agreement with noisy data, i.e., \(E_{\text{incompr}} = 7.25 \cdot 10^{-2}\).
To further assess the agreement between the true and discovered models, their displacement and reaction force predictions can be computed for the considered experimental setup. As a representative example, \cref{fig:displacement_reaction_force_Demiray} shows the displacements and reaction forces predicted by the true and discovered Demiray model for the highest noise level.

Overall, our results demonstrate the effectiveness of both the supervised and unsupervised \MF{} in accurately identifying material models across different noise levels. The findings highlight the resilience and adaptability of the method in handling noise while acknowledging the potential deviations that higher noise may introduce. The close relationship between the Neo-Hooke and Ogden models emphasizes the importance of considering model equivalency when analyzing material responses, particularly for small stretch and strain levels.

We finally emphasize the importance of fingerprint normalization and parameter rescaling in \MF{}, which \CHANGE{allows for accurately determining parameter magnitudes} and facilitates model discovery even when the order of magnitude of the true parameters is not present in the database.
\CHANGE{Due to the fingerprint normalization, a single fingerprint in the database is sufficient to cover all possible parameter choices of models that depend on a single parameter.}
\CHANGE{Moreover, the rescaling enables \MF{} to uncover material models without requiring the parameters in the database to correspond to their true physical magnitudes.}
For instance, the Blatz-Ko model is correctly discovered although the parameter $\theta_1 = 50.00$ of the benchmark model is far outside the range of parameters in the database, see \cref{tab:database_supervised,tab:database_unsupervised}.

\section{Conclusions}
\label{sec:conclusions}

\MF{} is a powerful and versatile technique for rapid material model discovery.
It simultaneously identifies both the interpretable functional form of a material model and its parameters
without requiring the solution of a complex optimization problem, thereby avoiding the risk of convergence to local optima.
By ensuring that only physically admissible models are included in the database, \MF{} inherently guarantees the discovery of physically meaningful models.
We demonstrated the effectiveness of \MF{} in the supervised setting using uniaxial tension and simple shear experiments, as well as in the unsupervised setting with heterogeneous deformation data.
However, the underlying framework is general and can be applied to any experimental setup.

\MF{} offers substantial potential for future extensions.
While this study focused on hyperelastic materials, the framework is inherently general and can be extended to a broader class of material behaviors, including dissipative materials.
A critical component of \MF{} is the construction of a high-quality database.
Our initial implementation employed relatively small databases as a proof of concept.
To enable broad applicability and reproducibility, the underlying experiments used to generate fingerprints should be standardized and designed for consistent implementation across different laboratories.
We envision that, in the future, a collaborative effort by experts in the field will lead to the development of a comprehensive, high-fidelity database containing a wide range of material models tailored to the most common experimental setups.
Although this requires a one-time investment of time and resources, the resulting database can be reused indefinitely, enabling instantaneous model discovery across diverse experiments.
Data compression techniques may be employed to reduce either the number of fingerprint vectors or the dimensionality of each fingerprint.
To further accelerate pattern recognition, the database can be partitioned into subgroups, each represented by a characteristic fingerprint.
This hierarchical structure allows for a coarse initial search across representative fingerprints, followed by a refined search within the most relevant subgroup.
Additionally, incorporating a sparsity-promoting regularization term into the pattern recognition algorithm may improve interpretability.
as an immediate next step, we aim to validate \MF{} experimentally.


\section*{Code and data availability}

Code and data \CHANGE{related to this work} are publicly available on Zenodo, see \cite{flaschel_supplementary_2025-1}, and on GitHub at
\begin{center}
    \url{https://github.com/Material-Fingerprinting/material-fingerprinting-hyperelasticity} .
\end{center}

\section*{Acknowledgments}

The authors acknowledge support from the European Research Council (ERC) Grant 101141626 DISCOVER. Funded by the European Union. Views and opinions expressed are, however, those of the authors only and do not necessarily reflect those of the European Union or the European Research Council
Executive Agency. Neither the European Union nor the granting authority can be held responsible for them.
The authors utilized ChatGPT, Perplexity and HAWKI, a large language model interface provided by Friedrich-Alexander-Universität Erlangen-Nürnberg, to enhance the writing style in certain sections of the manuscript.
After using these tools, the authors reviewed and edited the content as needed and take full responsibility for the content of the publication.


\appendix

\section{Uniaxial tension and simple shear experiments}
\label{sec:uniaxial_tension_simple_shear}

We derive expressions for the first Piola-Kirchhoff stress tensor \( \bfP \) from the strain energy density \( W \) for incompressible materials, as considered in the supervised \MF{}.
A Lagrange multiplier $p$ is added to the strain energy density function to impose the incompressibility constraint \( J=\det \bfF = 1 \).
\CHANGE{We distinguish between invariant-based models that depend on the first and second invariant of the right Cauchy-Green tensor $\bfC=\bfF^T \cdot \bfF$, that is, $I_1=\tr(\bfC)$ and $I_2=\frac{1}{2}[\tr(\bfC)^2 - \tr(\bfC^2)]$, and models that depend on the principal stretches $\lambda_1, \ \lambda_2, \lambda_3$, that is, the square roots of the eigenvalues of $\bfC$.}
For the invariant-based models, we have
\be
W = \tilde{W}(I_1, I_2) \CHANGE{- p [J-1]}.
\ee
Consequently, the first Piola-Kirchhoff stress tensor is given by
\be
\bfP = \frac{\partial \tilde{W}}{\partial \bfF} - p \, \bfF^{-T}.
\ee
For uniaxial tension, the deformation gradient \( \bfF_{\text{UT}} \) is given in \cref{eq: deformation gradient ut ss}. By setting \( F_{11} = \lambda \), the invariants become
\be
I_1 = \lambda^2 + 2\lambda^{-1}, \quad
I_2 = 2\lambda + \lambda^{-2}
\ee

The first Piola-Kirchhoff stress component in the loading direction is
\be
\label{eq:P11}
P_{11} = \frac{\partial \tilde{W}}{\partial \lambda} - p F_{11}
= \frac{\partial \tilde{W}}{\partial I_1} \frac{\partial I_1}{\partial \lambda}
+ \frac{\partial \tilde{W}}{\partial I_2} \frac{\partial I_2}{\partial \lambda} - p F_{11},
\ee
where
\be
\frac{\partial I_1}{\partial \lambda} = 2\lambda - 2\lambda^{-2}, \quad
\frac{\partial I_2}{\partial \lambda} = 2 - 2\lambda^{-3}.
\ee

The pressure term \( p \) is determined by imposing the plane stress conditions \( P_{22} = P_{33} = 0 \).

Simple shear is characterized by the deformation gradient \( \bfF_{\text{SS}} \) given in \cref{eq: deformation gradient ut ss}. By setting \( F_{12} = \gamma \), we obtain the invariants
\be
I_1 = I_2 = 3 + \gamma^2,
\ee
and the shear stress component becomes
\be
\label{eq:P12}
P_{12} = \frac{\partial \tilde{W}}{\partial \gamma} = \frac{\partial \tilde{W}}{\partial I_1} \frac{\partial I_1}{\partial \gamma}
+ \frac{\partial \tilde{W}}{\partial I_2} \frac{\partial I_2}{\partial \gamma}
= 2\gamma \left[ \frac{\partial \tilde{W}}{\partial I_1} + \frac{\partial \tilde{W}}{\partial I_2} \right].
\ee
We note that shear stresses in simple shear are independent of the pressure \( p \). 

For the Ogden model, which depands on the principal stretches, we have
\be
W = \tilde{W}(\lambda_1, \lambda_2, \lambda_3) - p [J-1].
\ee
The principal stretches are computed under uniaxial tension as
\be
\lambda_1 = \lambda, \quad
\lambda_2 = \lambda_3 = \frac{1}{\sqrt{\lambda}},
\ee
and under simple shear as
\be
\lambda_1 = \sqrt{\left[1 + \frac{1}{2}\gamma^2\right] - \sqrt{\left[1 + \frac{1}{2}\gamma^2\right]^2 - 1}},
\quad
\lambda_2 = 1,
\quad
\lambda_3 = \sqrt{\left[1 + \frac{1}{2}\gamma^2\right] + \sqrt{\left[1 + \frac{1}{2}\gamma^2\right]^2 - 1}}.
\ee
The stress components for the Ogden model can then be computed using formulas \cref{eq:P11,eq:P12}.

The remaining derivatives for the models specified in \cref{tab:database_supervised} are:
\begin{itemize}
\item {Blatz-Ko:} 
$
\frac{\partial \tilde{W}}{\partial I_2} = \theta_1
$,
\item {Demiray:} 
$
\frac{\partial \tilde{W}}{\partial I_1} = \theta_2 \alpha_2 \exp\left[\alpha_2 [I_1 - 3]\right]
$,
\item {Gent:} 
$
\frac{\partial \tilde{W}}{\partial I_1} = \frac{\theta_3 \alpha_3}{1 - \alpha_3 [I_1 - 3]}
$,
\item {Holzapfel:} 
$
\frac{\partial \tilde{W}}{\partial I_1} = 2 \theta_4 \alpha_4 [I_1 - 3] \exp\left[\alpha_4 [I_1 - 3]^2\right]
$,
\item {Mooney-Rivlin:} 
$
\frac{\partial \tilde{W}}{\partial I_1} = \theta_5, \quad 
\frac{\partial \tilde{W}}{\partial I_2} = \theta_1
$,
\item {Neo-Hooke:} 
$
\frac{\partial \tilde{W}}{\partial I_1} = \theta_5
$,
\item {Ogden:} 
$
\frac{\partial \tilde{W}}{\partial \lambda_i} = \theta_6 \alpha_c \lambda_i^{\alpha_c - 1}, 
\quad i \in \{1, 2, 3\}
$.
\end{itemize}

\section{Fingerprint database generation}

\subsection{Supervised \MF{}}
\label{sec:database_supervised}

\begin{table}[h!]
\caption{
Material models considered during database generation in the supervised setting.
}
\label{tab:database_supervised}
\centering
\begin{tabular}{|l|l|l|r|}
\hline
Models\vphantom{$\int_a^b$} & Strain energy density $W$ & Parameters ranges & \# Fingerprints \\ \hline

\rowcolor{BlatzKo!12}
\multicolumn{1}{|l|}{Blatz-Ko\vphantom{$\int_a^b$}} & $\theta_1[I_2 - 3] - p [J-1]$ & $\theta_1 = 1.0$ & 1 \\ 

\rowcolor{Demiray!12}
\multicolumn{1}{|l|}{Demiray\vphantom{$\int_a^b$}} & $\theta_2 [\exp(\alpha_2  [I_1 - 3]) - 1] - p [J-1]$ & $\theta_2 = 1.0$, $\alpha_2 \in [0.1, 10.0]$ & 100 \\ 

\rowcolor{Gent!12}
\multicolumn{1}{|l|}{Gent\vphantom{$\int_a^b$}} & $- \theta_3 [\ln(1 - \alpha_3 [I_1 - 3])] - p [J-1]$ & $\theta_3 = 1.0$, $\alpha_3 \in [0.1, 1.0]$ & 100 \\ 

\rowcolor{Holzapfel!12}
\multicolumn{1}{|l|}{Holzapfel\vphantom{$\int_a^b$}} & $\theta_4 [\exp(\alpha_4  [I_1 - 3]^2) - 1] - p [J-1]$ & $\theta_4 = 1.0$, $\alpha_4 \in [0.1, 10.0]$ & 100 \\ 

\rowcolor{MooneyRivlin!12}
\multicolumn{1}{|l|}{Mooney-Rivlin\vphantom{$\int_a^b$}} & $\theta_5 \left[I_1 - 3\right] + \theta_1\left[I_2 - 3\right] - p [J-1]$ & $\theta_1 = 1.0$, $\theta_5 \in [0.1, 10.0]$ & 100 \\

\rowcolor{NeoHooke!12}
\multicolumn{1}{|l|}{Neo-Hooke\vphantom{$\int_a^b$}} & $\theta_5 [I_1 - 3] - p [J-1]$ & $\theta_5 = 1.0$ & 1 \\ 

\rowcolor{Ogden!12}
\multicolumn{1}{|l|}{Ogden\vphantom{$\int_a^b$}} & $\theta_6 [\lambda_1^{\alpha_6} + \lambda_2^{\alpha_6} + \lambda_3^{\alpha_6} - 3] - p [J-1]$ & $\theta_6 = 1.0$, $\alpha_6 \in [0.1, 10.0]$ & 100 \\ 

\hline
\multicolumn{3}{|c|}{\vphantom{$\int_a^b$}} & $n_d=$ 502 \\
\hline
\end{tabular}
\end{table}

\begin{figure}
    \centering
    \includegraphics[width=0.8\textwidth]{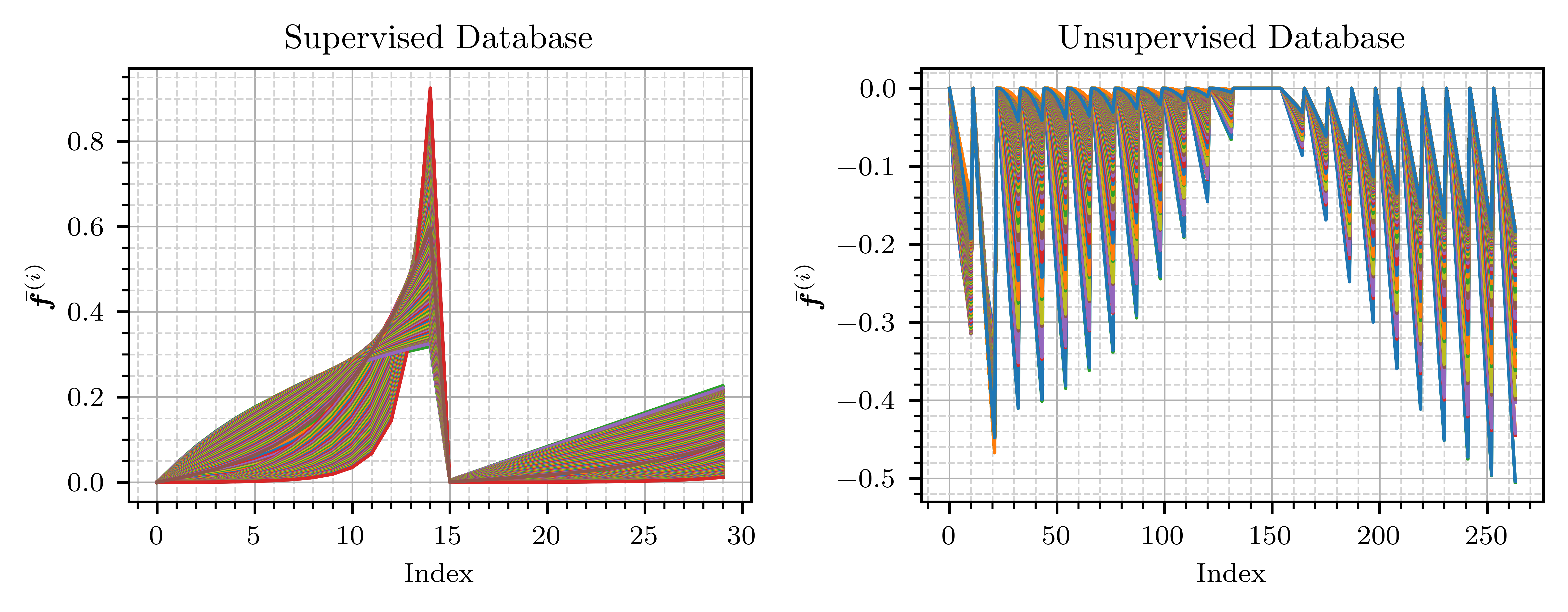}
    \caption{Illustration of all normalized fingerprints in the supervised (left) and unsupervised (right) database.}
    \label{fig:fingerprints}
\end{figure}

In this work, we consider incompressible material behavior in the supervised case of \MF{}. To generate the database, we consider classical incompressible hyperelastic material models listed in~\cref{tab:database_supervised}. These include Blatz-Ko, Demiray, Gent, Holzapfel, Mooney–Rivlin, Neo-Hooke, and Ogden models. For each model, material parameters are sampled using 100 equidistant values over the specified ranges, except for the Blatz-Ko and Neo-Hooke models, which are used with one fixed homogeneity parameter, resulting in a single but sufficient fingerprint each, see \cref{sec: Fingerprint database generation} for the homogeneity property. In total, we generate 502 fingerprints in this setting, though the database can readily accommodate additional entries. Fingerprints are obtained from simulated uniaxial tension and simple shear experiments, as detailed in \cref{sec:uniaxial_tension_simple_shear}. For uniaxial tension, we apply $n_{\text{UT}} = 15$ equidistant stretch values in the range $\lambda \in [1.0, 1.5]$, and for simple shear, $n_{\text{SS}} = 15$ equidistant shear values in the range $\gamma \in [0.0, 0.5]$\CHANGE{, yielding a total number of $n_f=30$ fingerprint components}. \cref{fig:fingerprints} illustrates the normalized fingerprints stored in the database.

\subsection{Unsupervised \MF{}}
\label{sec:database_unsupervised}

In the unsupervised setting, we assume compressible material behavior. We consider a set of the models listed in~\cref{tab:database_unsupervised}. Specifically, we include the compressible Blatz-Ko, Demiray, Gent, Mooney–Rivlin, and Neo-Hooke models, formulated in terms of the isochoric invariants $\bar I_1 = J^{-2/3}I_1$ and $\bar I_2=J^{-4/3}I_2$ and a volumetric penalty term $\theta_0 [J - 1]^2$ with $\theta_0 > 0$. For each model, parameters are sampled using 100 equidistant values across the specified ranges, resulting in a total of 30,200 fingerprints. These fingerprints are generated by simulating a plate with a hole subjected to biaxial tension over $n_t = 10$ load steps, see \cref{fig:plate}. Displacements are recorded at $n_u=11$ locations around the hole, and the net reaction forces at the boundary are computed. \CHANGE{Including an additional zeroth load step at which all displacements and forces are zero, this gives a total of $n_f=264$ fingerprint components.} The database was generated in less than one hour on a laptop. The normalized fingerprints are illustrated in \cref{fig:fingerprints}. In the future, larger databases can be constructed, for example by leveraging computing clusters and parallel processing. Further, data compression methods may be explored \citep{mcgivney_svd_2014}.

\begin{table}[h!]
\caption{
Material models considered during database generation in the unsupervised setting.
}
\label{tab:database_unsupervised}
\centering
\resizebox{\textwidth}{!}{
\begin{tabular}{|l|l|l|l|r|}
\hline
Models\vphantom{$\int_a^b$} & Strain energy density $W$ & \multicolumn{2}{c|}{Parameters ranges} & \# Fingerprints \\ \hline

\rowcolor{BlatzKo!12}
\multicolumn{1}{|l|}{Blatz-Ko\vphantom{$\int_a^b$}} & $\theta_1 [\bar I_2 - 3] + \theta_0 [J-1]^2$ & $\theta_1 = 1.0$ & $\theta_0 \in [0.1, 10.0]$ & 100\\ 

\rowcolor{Demiray!12}
\multicolumn{1}{|l|}{Demiray\vphantom{$\int_a^b$}} & $\theta_2 [\exp(\alpha_2  [\bar I_1 - 3]) - 1] + \theta_0 [J-1]^2$ & $\theta_2 = 1.0$, $\alpha_2 \in [0.1, 10.0]$ & $\theta_0 \in [0.1, 10.0]$ & 10,000\\ 

\rowcolor{Gent!12}
\multicolumn{1}{|l|}{Gent\vphantom{$\int_a^b$}} & $- \theta_3 [\ln(1 - \alpha_3 [\bar I_1 - 3])] + \theta_0 [J-1]^2$ & $\theta_3 = 1.0$, $\alpha_3 \in [0.1, 1.0]$ & $\theta_0 \in [0.1, 10.0]$ & 10,000\\ 


\rowcolor{MooneyRivlin!12}
\multicolumn{1}{|l|}{Mooney-Rivlin\vphantom{$\int_a^b$}} & $\theta_5 [\bar I_1 - 3] + \theta_1[\bar I_2 - 3] + \theta_0 [J-1]^2$ & $\theta_5 = 1.0$, $\theta_1 \in [0.1, 10.0]$ & $\theta_0 \in [0.1, 10.0]$ & 10,000\\

\rowcolor{NeoHooke!12}
\multicolumn{1}{|l|}{Neo-Hooke\vphantom{$\int_a^b$}} & $\theta_5 [\bar I_1 - 3] + \theta_0 [J-1]^2$ & $\theta_5 = 1.0$ & $\theta_0 \in [0.1, 10.0]$ & 100\\ 

\hline
\multicolumn{4}{|c|}{\vphantom{$\int_a^b$}} & $n_d=$ 30,200 \\
\hline
\end{tabular}
}
\end{table}

\section{\MF{} using databases that exclude the ground truth model}
\label{sec:database_corrupted}

\begin{figure}[h!]
\centering
\begin{subfigure}{0.7\textwidth}
\centering
\includegraphics[width=\linewidth]{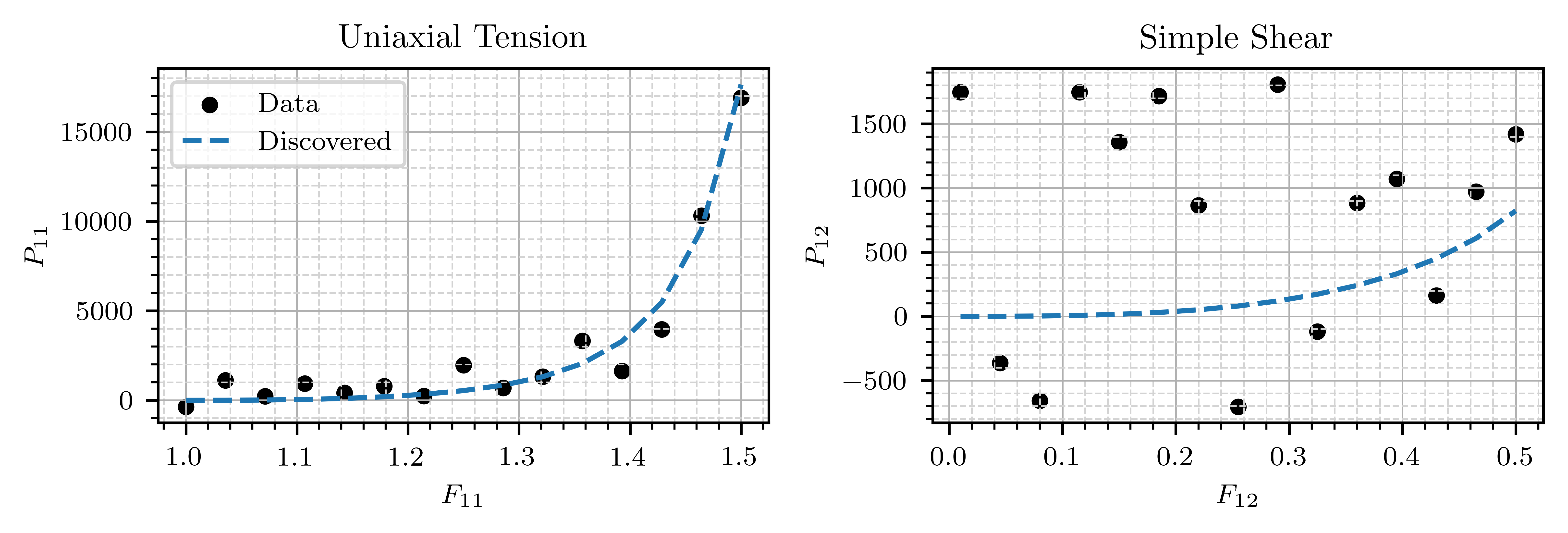}
\caption{Artificially generated data using the Demiray model with 5\% noise, and the discovered Holzapfel model using an intentionally corrupted database that excludes the ground truth model.}
\end{subfigure}\\%
\begin{subfigure}{0.68\textwidth}
\centering
\hspace*{0.1cm}
\includegraphics[width=\linewidth]{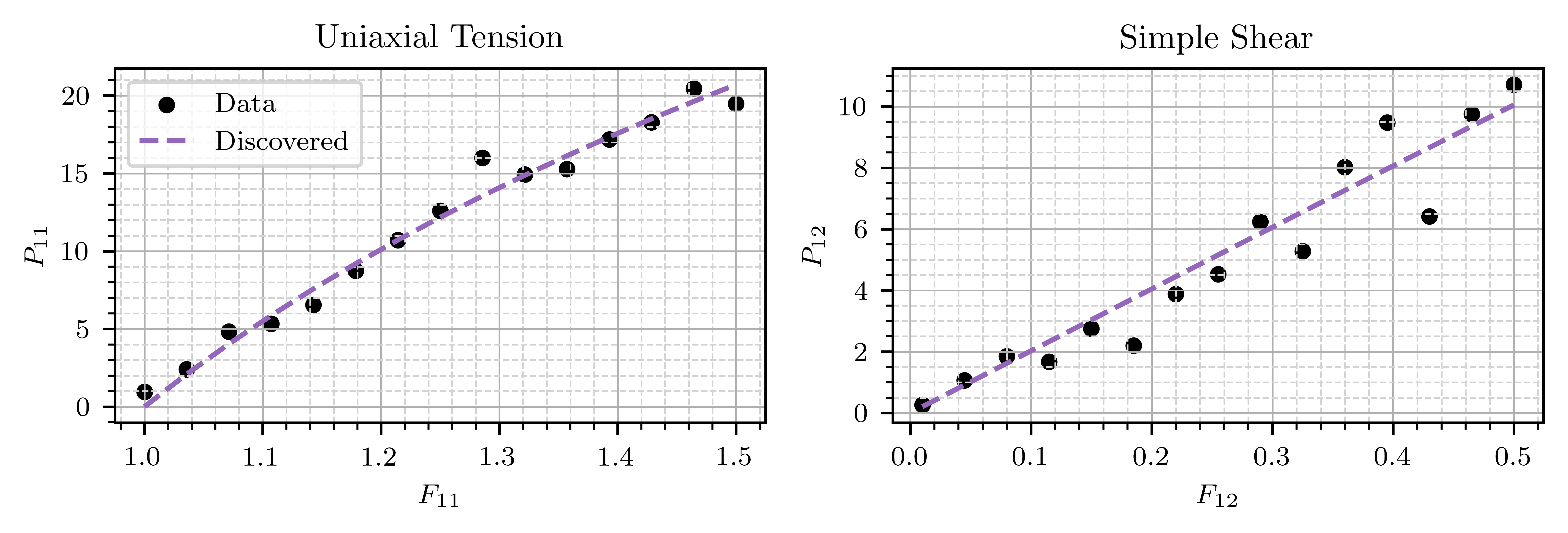}
\caption{Artificially generated data using the Neo-Hooke model with 5\% noise, and the discovered Ogden model using an intentionally corrupted database that excludes the ground truth model.}
\end{subfigure}\\%
\caption{\CHANGETWO{Stress response of the discovered models using intentionally corrupted databases that exclude the ground truth model in comparison to the uniaxial tension (left) and simple shear (right) data with the highest noise level.}
}
\label{fig:stress_strain_corrupted}
\end{figure}

\CHANGETWO{To verify the generalizability of \MF{}, we apply it to the supervised benchmark data using intentionally corrupted databases that exclude the ground truth model in each case.
As shown in \cref{tab:benchmarks_supervised_corrupted}, \MF{} with the corrupted databases discovers surrogate models that are not equal to but approximate the ground truth models.
For example, the Demiray model is approximated by the Holzapfel model and for the Neo-Hooke model, an alternative formulation in form of the Ogden model is discovered.
As expected, the error $E_{\text{incompr}}$ increases due to the deviation between the true and discovered strain energy density.
However, as the surrogate models mimic the ground truth model, at least within the regime of stretches and strains available in the input data, the mismatch between the measured and predicted stresses does not change significantly.
This is confirmed by the coefficients of determination $R^2$, which are comparable to those obtained using the complete database in \cref{tab:benchmarks_supervised}.
Plotting the discovered model responses in comparison to the measurements in \cref{fig:stress_strain_corrupted} shows that the surrogate models are capable of adequately describing the given data.}

\begin{table}[h!]
\caption{\CHANGETWO{Strain energy density functions discovered through supervised \MF{} using intentionally corrupted databases that exclude the ground truth model. Different colors indicate different material models in the database.}}
\label{tab:benchmarks_supervised_corrupted}
\centering
\begin{tabular}{|ll|l|r|r|}
\hline
\multicolumn{2}{|l|}{Benchmarks\vphantom{$\int_a^b$}} & \multicolumn{1}{c|}{Strain energy density $W$} & \multicolumn{1}{c|}{$E_{\text{incompr}}$} & \multicolumn{1}{c|}{$R^2$} \\ \hline

\rowcolor{BlatzKo!12}
\multicolumn{1}{|l|}{Blatz-Ko\vphantom{$\int_a^b$}} & Truth & $50.00 [I_2 - 3] - p [J-1]$ & -- & -- \\ \cline{2-5}
\rowcolor{MooneyRivlin!12}
\multicolumn{1}{|l|}{\vphantom{$\int_a^b$}} & $0\%$ Noise & $4.44 [I_1 - 3] + 44.40 [I_2 - 3] - p [J-1]$ & \num{2.34e-02} & 0.9990 \\ \cline{2-5}
\rowcolor{MooneyRivlin!12}
\multicolumn{1}{|l|}{\vphantom{$\int_a^b$}} & $1\%$ Noise & $4.47 [I_1 - 3] + 44.71 [I_2 - 3] - p [J-1]$ & \num{1.92e-02} & 0.9988 \\ \cline{2-5}
\rowcolor{MooneyRivlin!12}
\multicolumn{1}{|l|}{\vphantom{$\int_a^b$}} & $5\%$ Noise & $4.37 [I_1 - 3] + 43.72 [I_2 - 3] - p [J-1]$ & \num{3.62e-02}  & 0.9981 \\ \hline

\rowcolor{Demiray!12}
\multicolumn{1}{|l|}{Demiray\vphantom{$\int_a^b$}} & Truth & $10.00 \left[\exp(8.00 [I_1 - 3]) - 1\right] - p [J-1]$ & -- & -- \\ \cline{2-5}
\rowcolor{Holzapfel!12}
\multicolumn{1}{|l|}{\vphantom{$\int_a^b$}} & $0\%$ Noise & $158.48 \left[\exp(6.00 [I_1 - 3]^2) - 1\right] - p [J-1]$ & \num{2.07e+00} & 0.9996 \\ \cline{2-5}
\rowcolor{Holzapfel!12}
\multicolumn{1}{|l|}{\vphantom{$\int_a^b$}} & $1\%$ Noise & $158.81 \left[\exp(6.00 [I_1 - 3]^2) - 1\right] - p [J-1]$ & \num{2.07e+00} & 0.9996 \\ \cline{2-5}
\rowcolor{Holzapfel!12}
\multicolumn{1}{|l|}{\vphantom{$\int_a^b$}} & $5\%$ Noise &  $222.68 \left[\exp(5.30 [I_1 - 3]^2) - 1\right] - p [J-1]$ & \num{8.86e-01} & 0.9974 \\ \hline

\rowcolor{MooneyRivlin!12}
\multicolumn{1}{|l|}{Mooney-Rivlin\vphantom{$\int_a^b$}} & Truth & $10.00 [I_1 - 3] + 40.00 [I_2 - 3] - p [J-1]$ & -- & -- \\ \cline{2-5}
\rowcolor{Ogden!12}
\multicolumn{1}{|l|}{\vphantom{$\int_a^b$}} & $0\%$ Noise & $19306.72 [\lambda_1^{0.10} + \lambda_2^{0.10} + \lambda_3^{0.10} - 3] - p [J-1]$ & \num{5.60e-02} & 0.9957 \\ \cline{2-5}
\rowcolor{BlatzKo!12}
\multicolumn{1}{|l|}{\vphantom{$\int_a^b$}} & $1\%$ Noise & $52.69 [I_2 - 3] - p [J-1]$ & \num{5.34e-02} & 0.9955 \\ \cline{2-5}
\rowcolor{Ogden!12}
\multicolumn{1}{|l|}{\vphantom{$\int_a^b$}} & $5\%$ Noise & $19200.31 [\lambda_1^{0.10} + \lambda_2^{0.10} + \lambda_3^{0.10} - 3] - p [J-1]$ & \num{5.91e-02} & 0.9956 \\ \hline

\rowcolor{NeoHooke!12}
\multicolumn{1}{|l|}{Neo-Hooke\vphantom{$\int_a^b$}} & Truth & $10.00 [I_1 - 3] - p [J-1]$ & -- & -- \\ \cline{2-5}
\rowcolor{Ogden!12}
\multicolumn{1}{|l|}{\vphantom{$\int_a^b$}} & $0\%$ Noise & $10.00 [\lambda_1^{2.00} + \lambda_2^{2.00} + \lambda_3^{2.00} - 3] - p [J-1]$ & \num{0.00} & 1.0000 \\ \cline{2-5}
\rowcolor{Ogden!12}
\multicolumn{1}{|l|}{\vphantom{$\int_a^b$}} & $1\%$ Noise & $9.97 [\lambda_1^{2.00} + \lambda_2^{2.00} + \lambda_3^{2.00} - 3] - p [J-1]$ & \num{3.14e-03} & 1.0000 \\ \cline{2-5}
\rowcolor{Ogden!12}
\multicolumn{1}{|l|}{\vphantom{$\int_a^b$}} & $5\%$ Noise & $12.50 [\lambda_1^{1.80} + \lambda_2^{1.80} + \lambda_3^{1.80} - 3] - p [J-1]$ & \num{1.11e-02} & 0.9996 \\ \hline

\rowcolor{Ogden!12}
\multicolumn{1}{|l|}{Ogden\vphantom{$\int_a^b$}} & Truth & $5.00[\lambda_1^{8.00} + \lambda_2^{8.00} + \lambda_3^{8.00} - 3] - p [J-1]$ & -- & -- \\ \cline{2-5}
\rowcolor{Demiray!12}
\multicolumn{1}{|l|}{\vphantom{$\int_a^b$}} & $0\%$ Noise & $46.61 \left[\exp(2.10 [I_1 - 3]) - 1\right] - p [J-1]$ & \num{1.95e-01} & 0.9942 \\ \cline{2-5}
\rowcolor{Demiray!12}
\multicolumn{1}{|l|}{\vphantom{$\int_a^b$}} & $1\%$ Noise & $46.55 \left[\exp(2.10 [I_1 - 3]) - 1\right] - p [J-1]$ & \num{1.94e-01} & 0.9942 \\ \cline{2-5}
\rowcolor{Demiray!12}
\multicolumn{1}{|l|}{\vphantom{$\int_a^b$}} & $5\%$ Noise & $53.43 \left[\exp(2.00 [I_1 - 3]) - 1\right] - p [J-1]$ & \num{2.62e-01} & 0.9904 \\ \hline

\end{tabular}
\end{table}

\bibliographystyle{elsarticle-harv}
\bibliography{ALL}

\end{document}